\documentclass[9pt,shortpaper,twoside,web]{ieeecolor}

\usepackage{generic}

\usepackage{amsmath}
\usepackage{amssymb}
\usepackage{amsfonts}
\usepackage{mathtools}
\usepackage{bm}
\usepackage{mathrsfs}

\usepackage{tikz}
\usepackage{graphicx}
\usetikzlibrary{
    arrows.meta,
    positioning,
    shapes.geometric,
    calc
}

\usepackage{graphicx}
\usepackage{booktabs}
\usepackage{multirow}
\usepackage{float}

\usepackage{algorithm}
\usepackage{algorithmic}

\usepackage{textcomp}
\usepackage{xcolor}
\usepackage{url}
\usepackage{etoolbox}

\usepackage{tikz}
\usetikzlibrary{
    arrows.meta,
    positioning,
    shapes.geometric,
    calc
}

\usepackage{cite}

\definecolor{IEEElinkblue}{RGB}{0,85,190}

\usepackage[
    colorlinks=true,
    linkcolor=IEEElinkblue,
    citecolor=IEEElinkblue,
    urlcolor=IEEElinkblue,
    filecolor=IEEElinkblue,
    pdfborder={0 0 0},
    bookmarksnumbered=true,
    bookmarksopen=true
]{hyperref}

\AtBeginDocument{%
    \hypersetup{
        colorlinks=true,
        linkcolor=IEEElinkblue,
        citecolor=IEEElinkblue,
        urlcolor=IEEElinkblue
    }
}

\usepackage{etoolbox}

\AtBeginDocument{%
    \setlength{\abovedisplayskip}{1pt}%
    \setlength{\belowdisplayskip}{1pt}%
    \setlength{\abovedisplayshortskip}{0pt}%
    \setlength{\belowdisplayshortskip}{0pt}%
    \setlength{\jot}{0pt}%
}

\usepackage{titlesec}

\titlespacing*{\section}
{0pt}{5pt plus 1pt minus 1pt}{2pt}

\titlespacing*{\subsection}
{0pt}{4pt plus 1pt minus 1pt}{1.5pt}

\titlespacing*{\subsubsection}
{0pt}{3pt plus 1pt minus 1pt}{1pt}

\def\BibTeX{{\rm B\kern-.05em{\sc i\kern-.025em b}\kern-.08em
    T\kern-.1667em\lower.7ex\hbox{E}\kern-.125emX}}

\newtheorem{theorem}{Theorem}
\newtheorem{lemma}{Lemma}
\newtheorem{proposition}{Proposition}

\newtheorem{assumption}{Assumption}
\newtheorem{definition}{Definition}
\newtheorem{remark}{Remark}

\newcommand{\bx}{\bm{x}}
\newcommand{\bu}{\bm{u}}
\newcommand{\be}{\bm{e}}

\newcommand{\bfv}{\bm{f}}
\newcommand{\bg}{\bm{g}}
\newcommand{\bh}{\bm{h}}

\newcommand{\bW}{\bm{W}}

\newcommand{\bphi}{\bm{\phi}}

\newcommand{\bsigma}{\bm{\sigma}}

\newcommand{\bOmega}{\bm{\Omega}}
\newcommand{\bGamma}{\bm{\Gamma}}

\newcommand{\bR}{\bm{R}}
\newcommand{\bP}{\bm{P}}

\begin{document}


\title{
Inverse-Optimal Fixed-Time Integral Reinforcement Learning
for Unknown Nonlinear Leader-Follower Multi-Agent Systems
}


\author{
Tien Dat Vu and Minh Doan
\thanks{
T. D. Vu and M. Doan are with the Faculty of Mechanical
Engineering, Ho Chi Minh City University of Technology
(HCMUT), Vietnam National University Ho Chi Minh City
(VNU-HCM), Ho Chi Minh City, Vietnam
(e-mail: dat.vuv@hcmut.edu.vn;
minh.doan@hcmut.edu.vn).
}
\thanks{
Corresponding author: M. Doan
(e-mail: minh.doan@hcmut.edu.vn).
}
}


\maketitle

\begin{abstract}
This paper develops a distributed inverse-optimal fixed-time integral
reinforcement learning framework for unknown nonlinear leader-follower
multi-agent systems. The proposed method first reconstructs the unknown
agent dynamics from finite integral data. Based on the identified dynamics, a distributed
inverse-optimal formulation is constructed by prescribing a desired
fixed-time value function and recovering a compatible running cost.
A critic-only integral reinforcement learning architecture is then
developed using current integral data and finitely informative
experience-replay data, thereby avoiding the persistent-excitation
requirement. Practical fixed-time convergence of the critic-weight
errors is established in the presence of identification and
function-approximation residuals. Finally, a network-level composite
Lyapunov analysis proves practical fixed-time leader-following
consensus of the true unknown nonlinear multi-agent system.
Numerical simulations are provided to validate the developed theory.
\end{abstract}


\section{Introduction}
\label{sec:introduction}

Distributed optimal control of nonlinear leader--follower
multi-agent systems (MASs) requires each follower to
coordinate with its neighbors while optimizing a local
performance objective under coupled error dynamics
\cite{Shi2020DataBasedMAS,Liu2026DistributedOptimal,
Cai2022CooperativeControl,Nagahara2024MultiAgent}.
This problem becomes particularly challenging when
the agent dynamics are unknown and both learning and
formation errors must converge within an
initial-condition-independent time bound
\cite{Wang2024FixedTimeFormation,Yue2025FixedTimeMAS}.
Moreover, conventional optimal control relies on
Hamilton--Jacobi--Bellman (HJB) equations, whose
solution becomes increasingly difficult for nonlinear,
interconnected systems
\cite{Vamvoudakis2010,Shi2020DataBasedMAS,
Lian2024IntegralInverseRL,Wei2025IterativeADP}.

Inverse-optimal control (IOC)~\cite[Sec.~8.4, pp.~524--525]{Haddad2008NonlinearControl} offers an alternative by
constructing a performance objective from a prescribed
stabilizing feedback, while fixed-time Lyapunov design
provides an initial-condition-independent convergence
bound \cite{Jimenez2017Predefined,Polyakov2012,
Lian2024IntegralInverseRL}.
For unknown dynamics, integral concurrent learning
enables identification from finite trajectory data
without state differentiation or persistent excitation
\cite{Parikh2019ICL,Chowdhary2010CL,
Kamalapurkar2018RL}.
Integral reinforcement learning (IRL) further
constructs Bellman equations from trajectory
integrals, allowing value-function learning without
explicit drift
\cite{Modares2014,Lian2024IntegralInverseRL,
Kamalapurkar2018RL}.
These mechanisms motivate combining inverse-optimal
synthesis, integral identification, and fixed-time
critic learning within a distributed framework
\cite{Kokolakis2023FixedTime,Kokolakis2025Predefined,
Parikh2019ICL}.

Several closely related studies have addressed
individual aspects of this problem.
Kokolakis et al. developed fixed-time and
predefined-time critic-only reinforcement learning
for optimal feedback stabilization, using
non-Lipschitz learning and experience replay
\cite{Kokolakis2023FixedTime,Kokolakis2025Predefined}.
However, these formulations do not directly
address distributed leader--follower coordination
with interconnected unknown agent dynamics
\cite{Kokolakis2023FixedTime,Kokolakis2025Predefined}.
Gong et al. investigated fixed-time secure formation
control for second-order MASs under false-data
injection attacks through a graphical zero-sum
game and coupled Hamilton--Jacobi--Isaacs equations
\cite{Gong2025}.
Their setting differs from general unknown
nonlinear leader--follower systems requiring
integral identification and inverse-optimal
feedback construction \cite{Gong2025}.
Meanwhile, data-driven distributed optimal
control and adaptive leader--follower methods
have addressed nonlinear dynamics through
adaptive dynamic programming or online HJB
approximation, without establishing the same
joint practical fixed-time learning and
coordination guarantee
\cite{Shi2020DataBasedMAS,Liu2026DistributedOptimal}.

Consequently, the remaining challenge is to
jointly reconstruct unknown nonlinear agent
dynamics, construct a distributed inverse-optimal
fixed-time objective, and learn the resulting
value functions from local integral data
\cite{Parikh2019ICL,Kokolakis2025Predefined,
Liu2026DistributedOptimal}.
In particular, a unified practical fixed-time
analysis accounting for identification residuals,
critic approximation errors, and interconnected
leader--follower dynamics remains insufficiently
developed
\cite{Gong2025,Wang2024FixedTimeFormation,
Kokolakis2025Predefined}.


The main contributions of this paper are fourfold.
First, a unified theoretical framework is established
that integrates data-driven identification,
inverse-optimal control, and integral reinforcement
learning for unknown nonlinear multi-agent systems.
Second, an inverse-optimal fixed-time control
formulation is developed to construct a suitable
running cost and feedback policy, ensuring
practical fixed-time stabilization despite
identification uncertainties.
Third, a novel two-power neural learning law
is proposed to guarantee practical fixed-time
convergence of the critic weights using current
and replayed integral data.
Fourth, a distributed practical fixed-time
control scheme is established for unknown
nonlinear leader--follower systems, with
explicit convergence-time bounds and residual
sets derived through composite Lyapunov analysis.

\noindent\textit{Notation:}
$\mathbb R^n$ and $\mathbb R^{n\times m}$
denote real vector and matrix spaces.
Bold symbols denote vectors or matrices,
whereas nonbold symbols denote scalars.
$\|\cdot\|$ denotes the Euclidean vector norm
or induced matrix $2$-norm, 
the infinity norm
is defined as
\(
\|\bx\|_\infty:=\max_j |x_j|
\).
$(\cdot)^\top$, $\otimes$, and $I_n$
denote transpose, Kronecker product, and
the identity matrix, respectively.
$\operatorname{col}(\cdot)$ and
$\operatorname{diag}(\cdot)$ denote
vertical stacking and diagonal assembly.
$\bP\succ0$ and $\bP\succeq0$ denote
positive definiteness and semidefiniteness,
respectively, for symmetric matrices.
$\lambda_{\min}(\bP)$ and
$\lambda_{\max}(\bP)$ denote its extreme
eigenvalues.
$\nabla V$ denotes the gradient of a
scalar function, and $C^1(\Omega)$ denotes
continuously differentiable functions on
$\Omega$. The notation \(O(\cdot)\) denotes the standard
Big-O asymptotic order. The symbols
\(\mathbf 0_n\) and \(\mathbf 1_n\) denote
the \(n\)-dimensional all-zero and all-one
column vectors, respectively.

\section{Preliminaries and Problem Formulation}
\label{sec:preliminaries}

\subsection{Preliminaries}

Consider a directed graph
\(\mathcal G=(\mathcal V,\mathcal E,\mathcal A)\), where
\(\mathcal V=\{1,\ldots,N\}\),
\(\mathcal E\subseteq\mathcal V\times\mathcal V\), and
\(\mathcal A=[a_{ij}]\in\mathbb R^{N\times N}\) is the adjacency
matrix. The weight \(a_{ij}>0\) means that follower \(i\) receives
information from follower \(j\), and \(a_{ij}=0\) otherwise. Define
\(\mathcal N_i:=\{j\in\mathcal V:(j,i)\in\mathcal E\}\),
\(\mathcal L:=\mathcal D-\mathcal A\),
\(\mathcal D:=\operatorname{diag}
\{\sum_{j=1}^{N}a_{ij}\}\),
\(\mathcal B:=\operatorname{diag}(b_1,\ldots,b_N)\), and
\(\mathcal H:=\mathcal L+\mathcal B\), where \(b_i\geq0\) denotes
the pinning gain from the leader to follower \(i\)~\cite{RenBeard2008,LiDuan2014}.

\begin{assumption}[Leader reachability]
\label{ass:graph}
The leader is globally reachable from the follower graph.
Equivalently, \(\mathcal H\) is nonsingular.
\end{assumption}

Under Assumption~\ref{ass:graph}, one has
\(\sigma_{\min}(\mathcal H)>0\), and therefore
\(\mathcal H^{-1}\) exists.

\begin{lemma}[Fixed-time practical convergence~\cite{Polyakov2012}]
\label{lem:fixed_time}
Let \(Y:[0,\infty)\rightarrow\mathbb R_{\geq0}\) be absolutely
continuous and suppose that
\(\dot Y\leq-aY^\mu-bY^\nu+\Delta\), where
\(a,b>0\), \(0<\mu<1<\nu\), and \(\Delta\geq0\).
For any \(\chi\in(0,1)\), define
\(\Omega_{Y,\chi}:=
\{Y\geq0:
aY^\mu+bY^\nu\leq\Delta/(1-\chi)\}\).
Then \(\Omega_{Y,\chi}\) is reached within the uniform time
\(T_{\max}\leq
1/[\chi a(1-\mu)]
+
1/[\chi b(\nu-1)]\),
which is independent of \(Y(0)\). If \(\Delta=0\), the origin is
fixed-time stable and
\(T_{\max}\leq
1/[a(1-\mu)]
+
1/[b(\nu-1)]\).
\end{lemma}

\subsection{Problem Formulation}
\label{subsec:problem_formulation}

Consider a leader--follower network consisting of one leader and \(N\)
nonlinear followers. The leader dynamic satisfies
\begin{equation}
    \dot{\bx}_0=\mathbf f_0(\bx_0),
    \qquad
    \bx_0\in\mathbb R^n,
    \label{eq:leader_dynamics}
\end{equation}
whereas the dynamic of follower \(i\) is described by
\begin{equation}
    \dot{\bx}_i
    =
    \mathbf f_i(\bx_i)
    +
    \mathbf g_i(\bx_i)\bu_i,
    \qquad
    i=1,\ldots,N,
    \label{eq:follower_dynamics}
\end{equation}
where \(\bx_i\in\mathbb R^n\), \(\bu_i\in\mathbb R^m\),
\(\mathbf f_i:\mathbb R^n\rightarrow\mathbb R^n\) is an unknown
nonlinear drift, and
\(\mathbf g_i:\mathbb R^n\rightarrow\mathbb R^{n\times m}\) is
known. 

\begin{assumption}[Compact operating domain]
\label{ass:compact_operating_domain}
Let \(\Omega_i\subset\mathbb R^n\) be a compact admissible operating
region containing the origin. The functions
\(\mathbf f_0(\bx_0)\), \(\mathbf f_i(\bx_i)\), and
\(\mathbf g_i(\bx_i)\) are assumed locally Lipschitz on their
respective operating regions, with
\(\mathbf f_i(\mathbf 0)=\mathbf 0\). Hence, for every admissible
control policy considered below, the corresponding closed-loop system
admits a unique forward solution on the prescribed operating domain.
\end{assumption}

\begin{definition}[Admissible control policy~\cite{Vamvoudakis2010}]
\label{def:admissible_policy}
A continuous feedback policy
\(
\bu_i:\Omega_i\to\mathcal U_i
\)
is admissible on
\(
\Omega_i
\)
if
\(
\bu_i(\mathbf0)=\mathbf0
\),
the corresponding closed-loop system is
asymptotically stable, and its associated
infinite-horizon cost (See~\eqref{eq:local_cost_function}) is finite.
\end{definition}

Let \(\mathbf h_i(t)\in\mathbb R^n\) denote
the desired relative offset of follower \(i\)
with respect to the leader
(for a time-invariant desired formation,
\(\mathbf h_i(t)\equiv\mathbf h_i\) is constant),
and define the individual leader-following
error
\(\tilde{\bx}_i:=\bx_i-\bx_0-\mathbf h_i(t)\).
The distributed local coordination error
is defined as
\begin{equation}
\mathbf e_i:=
\sum_{j\in\mathcal N_i}
a_{ij}(\tilde{\bx}_i-\tilde{\bx}_j)
+b_i\tilde{\bx}_i,
\label{eq:distributed_local_error}
\end{equation}

Equivalently,
\(\be_i=
\sum_{j\in\mathcal N_i}
a_{ij}[(\bx_i-\bh_i)-(\bx_j-\bh_j)]
+
b_i[(\bx_i-\bh_i)-\bx_0]\).
With
\(\tilde{\bx}:=\operatorname{col}
(\tilde{\bx}_1,\ldots,\tilde{\bx}_N)\) and
\(\be:=\operatorname{col}(\be_1,\ldots,\be_N)\), one has
\begin{equation}
    \be
    =
    (\mathcal H\otimes I_n)\tilde{\bx}.
    \label{eq:error_compact}
\end{equation}
Under Assumption~\ref{ass:graph},
\(\be=\mathbf 0\Longleftrightarrow\tilde{\bx}=\mathbf 0\), and
\(\|\tilde{\bx}\|
\leq
\|\mathcal H^{-1}\otimes I_n\|\,\|\be\|\).
Hence, leader-following consensus or formation tracking can be studied
through stabilization of the distributed error \(\be\).

Let $\Omega_{e_i}\subset\mathbb R^n$ denote a
compact admissible operating set for the local
neighborhood error $e_i$, containing the origin (\(\be_i=0\))
in its interior.


\begin{assumption}[Local error-state sufficiency]
\label{Local error-state sufficiency}
For each follower \(i\in\mathcal V\), the local
consensus error \(\be_i\) is assumed to be
a sufficient state for its error dynamics
on the operating region \(\Omega_{e_i}\).
Specifically, any two admissible network
configurations producing the same \(\be_i\),
under identical control inputs
\(\bu_i\) and \(\{\bu_j\}_{j\in\mathcal N_i}\),
and prescribed exogenous signals,
induce the same \(\dot{\be}_i\).
Consequently, the local error dynamics
admit a well-defined representation
in terms of \(\be_i\), \(\bu_i\),
and \(\{\bu_j\}_{j\in\mathcal N_i}\).
\end{assumption}

For compactness, define
\(
\mathbf F_i
:=
\mathbf f_i(\bx_i)
-
\mathbf f_0(\bx_0)
-
\dot{\bh}_i
\),
so that
\(
\dot{\tilde{\bx}}_i
=
\mathbf F_i
+
\mathbf g_i(\bx_i)\bu_i
\)
follows from
\eqref{eq:follower_dynamics}.
Differentiating
\eqref{eq:distributed_local_error}
yields
\(
\dot{\be}_i
=
\sum_{j\in\mathcal N_i}
a_{ij}
(\dot{\tilde{\bx}}_i-\dot{\tilde{\bx}}_j)
+
b_i\dot{\tilde{\bx}}_i
\).

Let
\(
\kappa_i
:=
b_i+\sum_{j\in\mathcal N_i}a_{ij}
\),
\(
\mathbf G_{ii}(\bx_i)
:=
\kappa_i\mathbf g_i(\bx_i)
\),
and
\(
\mathcal N_i
=
\{j_1,\ldots,j_{n_i^{\mathrm{nb}}}\}
\),
where
\(
n_i^{\mathrm{nb}}
:=
|\mathcal N_i|
\).
Define the stacked neighbor states and inputs as
\(
\bx_{\mathcal N_i}
:=
\operatorname{col}
(\bx_{j_1},\ldots,\bx_{j_{n_i^{\mathrm{nb}}}})
\)
and
\(
\bu_{\mathcal N_i}
:=
\operatorname{col}
(\bu_{j_1},\ldots,\bu_{j_{n_i^{\mathrm{nb}}}})
\),
respectively, together with
\(
\mathbf G_{i\mathcal N}
(\bx_{\mathcal N_i})
:=
\big[
-a_{ij_1}\mathbf g_{j_1}(\bx_{j_1}),
\ldots,
-a_{ij_{n_i^{\mathrm{nb}}}}
\mathbf g_{j_{n_i^{\mathrm{nb}}}}
(\bx_{j_{n_i^{\mathrm{nb}}}})
\big]
\).
Consequently,
\(
\mathbf G_{i\mathcal N}
(\bx_{\mathcal N_i})
\bu_{\mathcal N_i}
=
-\sum_{j\in\mathcal N_i}
a_{ij}\mathbf g_j(\bx_j)\bu_j
\).

The unknown local error drift is defined as
\(
\mathbf F_i^{e}
:=
\sum_{j\in\mathcal N_i}
a_{ij}(\mathbf F_i-\mathbf F_j)
+
b_i\mathbf F_i
\),
or equivalently,
\(
\mathbf F_i^{e}
=
\kappa_i
[
\mathbf f_i(\bx_i)
-
\mathbf f_0(\bx_0)
-
\dot{\bh}_i
]
-
\sum_{j\in\mathcal N_i}
a_{ij}
[
\mathbf f_j(\bx_j)
-
\mathbf f_0(\bx_0)
-
\dot{\bh}_j
]
\),
which incorporates the unknown nonlinear
follower dynamics, the leader dynamics,
and the desired-offset dynamics.

Consequently, the local distributed error dynamics can be written in
the compact form
\begin{equation}
    \dot{\be}_i
    =
    \mathbf F_i^{e}(\boldsymbol{\xi}_i)
    +
    \mathbf G_{ii}(\bx_i)\bu_i
    +
    \mathbf G_{i\mathcal N}
    (\bx_{\mathcal N_i})
    \bu_{\mathcal N_i},
    \label{eq:local_error_compact_dynamics}
\end{equation}
where
\(\boldsymbol{\xi}_i:=
\operatorname{col}
(\bx_i,\bx_{\mathcal N_i},\bx_0,
\dot{\bh}_i,\dot{\bh}_{\mathcal N_i})\), 
\(\dot{\bh}_{\mathcal N_i}:=
\operatorname{col}
(\dot{\bh}_{j_1},\ldots,
\dot{\bh}_{j_{n_i^{\mathrm{nb}}}})\).

\begin{remark}[Compact-domain limitation]
The compact set $\Omega_i$ is essential for the
subsequent identification, approximation, and
stability analysis, as commonly required in
neural Adaptive Dynamic Programming (ADP) \cite{Vamvoudakis2010,Modares2014}.
However, identifying such a set a priori is
nontrivial, and ensuring that trajectories
remain inside it may require a stabilizing
baseline or backup controller
\cite{Cohen2023SafeExploration,Rabiee2025BackupCBF}.
Even then, the size of the certified invariant
region may be difficult to characterize or
conservative \cite{Rabiee2025BackupCBF}.
Consequently, stability outside this region
is not guaranteed by the local analysis.
Control barrier functions offer a systematic
approach to enforcing forward invariance of
a prescribed admissible set
\cite{Ames2017CBF,Cohen2023SafeExploration}.
For analytical tractability, the present work
retains the compact operating domain as an
assumption.
\end{remark}

\section{Integral Composite Concurrent-Learning Identification}
\label{subsec:integral_identification_i}
inspired by the work of Wang et al~\cite{Wang2024FixedTimeFormation}. This Section is designed to identify the non-linear part within the dynamic equations, facilitating the inverse optimal control process described later.
Let
\(
\tau_{i,0}<\tau_{i,1}<\cdots
\)
denote the sampling instants used to construct integral identification
data for follower \(i\). Integrating the dynamics of follower \(i\) over
\(
[\tau_{i,\ell},\tau_{i,\ell+1}]
\)
and using the neural approximation
\(
\bfv_i(\bx_i)
=
\bW_{f_i}^{*\top}\bsigma_{f_i}(\bx_i)
+
\bm{\varepsilon}_{f_i}(\bx_i)
\)
gives
\begin{equation}
\begin{aligned}
    &\bx_i(\tau_{i,\ell+1})-\bx_i(\tau_{i,\ell})
    -
    \int_{\tau_{i,\ell}}^{\tau_{i,\ell+1}}
    \bg_i(\bx_i(\tau))\bu_i(\tau)\,d\tau
    \\
    &\qquad =
    \bW_{f_i}^{*\top}
    \int_{\tau_{i,\ell}}^{\tau_{i,\ell+1}}
    \bsigma_{f_i}(\bx_i(\tau))\,d\tau
    +
    \int_{\tau_{i,\ell}}^{\tau_{i,\ell+1}}
    \bm{\varepsilon}_{f_i}(\bx_i(\tau))\,d\tau .
\end{aligned}
\label{eq:integral_identification_i}
\end{equation}

For each interval, define
\(
\bm y_{f_i,\ell}
:=
\bx_i(\tau_{i,\ell+1})-\bx_i(\tau_{i,\ell})
-
\int_{\tau_{i,\ell}}^{\tau_{i,\ell+1}}
\bg_i(\bx_i(\tau))\bu_i(\tau)\,d\tau
\),
\(
\bm\phi_{f_i,\ell}
:=
\int_{\tau_{i,\ell}}^{\tau_{i,\ell+1}}
\bsigma_{f_i}(\bx_i(\tau))\,d\tau
\),
and
\(
\bm d_{f_i,\ell}
:=
\int_{\tau_{i,\ell}}^{\tau_{i,\ell+1}}
\bm{\varepsilon}_{f_i}(\bx_i(\tau))\,d\tau
\).
Then
\(
\bm y_{f_i,\ell}
=
\bW_{f_i}^{*\top}\bm\phi_{f_i,\ell}
+
\bm d_{f_i,\ell}
\).

The informative data are stored in the history stack
\(
\mathcal H_{f_i}
=
\{(\bm y_{f_i,\ell},\bm\phi_{f_i,\ell})\}_{\ell=0}^{M_{f_i}-1}
\),
with cumulative information matrix
\(
\bm{\mathcal S}_{f_i}
:=
\sum_{\ell=0}^{M_{f_i}-1}
\bm\phi_{f_i,\ell}\bm\phi_{f_i,\ell}^{\top}
\).
Whenever a new data segment is admitted,
\(
\bm{\mathcal S}_{f_i}
\)
is updated by
\(
\bm{\mathcal S}_{f_i}^{+}
=
\bm{\mathcal S}_{f_i}
+
\bm\phi_{f_i,\mathrm{new}}
\bm\phi_{f_i,\mathrm{new}}^{\top}
\).

\begin{assumption}[Finite excitation]
\label{ass:finite_excitation_i}
For follower \(i\), there exist a finite
\(M_{f_i}\geq N_{f_i}\) and a constant
\(\alpha_{f_i}>0\) such that the stored integral data satisfy
\(
\bm{\mathcal S}_{f_i}
=
\sum_{\ell=0}^{M_{f_i}-1}
\bm\phi_{f_i,\ell}\bm\phi_{f_i,\ell}^{\top}
\succeq
\alpha_{f_i}\bm I
\).
\end{assumption}

In addition to the stored data, define the current integral data over a
moving window of length \(T_{f_i}>0\) by
\(
\bm y_{f_i}(t)
:=
\bx_i(t)-\bx_i(t-T_{f_i})
-
\int_{t-T_{f_i}}^{t}
\bg_i(\bx_i(\tau))\bu_i(\tau)\,d\tau
\)
and
\(
\bm\phi_{f_i}(t)
:=
\int_{t-T_{f_i}}^{t}
\bsigma_{f_i}(\bx_i(\tau))\,d\tau
\).
From the follower dynamics and the neural approximation, these quantities
satisfy
\(
\bm y_{f_i}(t)
=
\bW_{f_i}^{*\top}\bm\phi_{f_i}(t)
+
\bm d_{f_i}(t)
\),
where
\(
\bm d_{f_i}(t)
:=
\int_{t-T_{f_i}}^{t}
\bm{\varepsilon}_{f_i}(\bx_i(\tau))\,d\tau
\).
Similarly, every stored pair satisfies
\(
\bm y_{f_i,\ell}
=
\bW_{f_i}^{*\top}\bm\phi_{f_i,\ell}
+
\bm d_{f_i,\ell}
\).

Using the current estimate \(\hat{\bW}_{f_i}(t)\), define the current and
stored integral prediction errors as
\(
\bm e_{f_i}(t)
:=
\bm y_{f_i}(t)
-
\hat{\bW}_{f_i}^{\top}(t)\bm\phi_{f_i}(t)
\)
and
\(
\bm e_{f_i,\ell}(t)
:=
\bm y_{f_i,\ell}
-
\hat{\bW}_{f_i}^{\top}(t)\bm\phi_{f_i,\ell}
\),
respectively. Consider the composite prediction-error objective
\(
J_{f_i}
:=
\frac{1}{2}\|\bm e_{f_i}(t)\|^2
+
\frac{k_{f_i}}{2}
\sum_{\ell=0}^{M_{f_i}-1}
\|\bm e_{f_i,\ell}(t)\|^2
\),
where \(k_{f_i}>0\). Since
\(
\nabla_{\hat{\bW}_{f_i}}J_{f_i}
=
-\bm\phi_{f_i}(t)\bm e_{f_i}^{\top}(t)
-
k_{f_i}
\sum_{\ell=0}^{M_{f_i}-1}
\bm\phi_{f_i,\ell}\bm e_{f_i,\ell}^{\top}(t)
\),
the gradient update
\(
\dot{\hat{\bW}}_{f_i}
=
-\bm{\Gamma}_{f_i}
\nabla_{\hat{\bW}_{f_i}}J_{f_i}
\)
gives
\begin{equation}
\dot{\hat{\bW}}_{f_i}
=
\bm{\Gamma}_{f_i}
\left[
    \bm\phi_{f_i}(t)\bm e_{f_i}^{\top}(t)
    +
    k_{f_i}
    \sum_{\ell=0}^{M_{f_i}-1}
    \bm\phi_{f_i,\ell}\bm e_{f_i,\ell}^{\top}(t)
\right].
\label{eq:icl_identifier_i}
\end{equation}
where
\(
\bm{\Gamma}_{f_i}
=
\bm{\Gamma}_{f_i}^{\top}\succ0
\).
Hence, the identifier combines the current moving-window information
with the informative historical data retained in the history stack.

Let
\(
\tilde{\bW}_{f_i}
:=
\hat{\bW}_{f_i}-\bW_{f_i}^{*}
\).
The above regression relations immediately yield
\(
\bm e_{f_i}
=
-\tilde{\bW}_{f_i}^{\top}\bm\phi_{f_i}
+
\bm d_{f_i}
\)
and
\(
\bm e_{f_i,\ell}
=
-\tilde{\bW}_{f_i}^{\top}\bm\phi_{f_i,\ell}
+
\bm d_{f_i,\ell}
\).
Accordingly, the identification dynamics are analyzed with
\(
V_{f_i}
:=
\frac{1}{2}
\operatorname{tr}
\left(
\tilde{\bW}_{f_i}^{\top}
\bm{\Gamma}_{f_i}^{-1}
\tilde{\bW}_{f_i}
\right)
\).


\begin{theorem}[\cite{Parikh2019IntegralCL,Pan2018CompositeLearning,Vu2026PredefinedIRL}]
\label{lem:identification_error_i}
Suppose that the preceding assumptions, including
Assumption~\ref{ass:finite_excitation_i}, hold and that the integral
approximation residuals are bounded.
Then the weight-estimation error generated by
\eqref{eq:icl_identifier_i} is uniformly ultimately bounded. In
particular, there exist constants \(c_{f_i}>0\) and
\(\Delta_{f_i}\geq0\) such that
\begin{equation}
    \dot V_{f_i}
    \leq
    -c_{f_i}V_{f_i}
    +
    \Delta_{f_i}.
    \label{eq:icl_weight_bound_i}
\end{equation}
If
\(
\bm{\varepsilon}_{f_i}\equiv\bm0
\),
then
\(
\Delta_{f_i}=0
\)
and
\(
\tilde{\bW}_{f_i}(t)\to\bm0
\).
\end{theorem}

The online reconstructed model
\(
\hat{\bfv}_i(\bx_i,t)
=
\hat{\bW}_{f_i}^{\top}(t)
\bsigma_{f_i}(\bx_i)
\)
is subsequently used in the inverse-optimal
construction and updated concurrently with the
closed-loop evolution.

\begin{lemma}[Bounded identification error]
\label{lem:bounded_identification_error_i}
Under the preceding assumptions and
Assumption~\ref{ass:finite_excitation_i},
the identification error is uniformly bounded on \(\Omega_i\). In
particular, there exists a finite constant
\(\bar\Delta_{f_i}>0\) such that
\begin{equation}
\left\|
\bfv_i(\bx_i)-\hat{\bfv}_i(\bx_i,t)
\right\|
\leq
\bar\Delta_{f_i},
\qquad
\forall\,\bx_i\in\Omega_i .
\label{eq:identification_error_bound_i}
\end{equation}
\end{lemma}

\begin{proof}
From
\(
\bfv_i(\bx_i)
=
\bW_{f_i}^{*\top}\bsigma_{f_i}(\bx_i)
+
\bm{\varepsilon}_{f_i}(\bx_i)
\)
and
\(
\hat{\bfv}_i(\bx_i,t)
=
\hat{\bW}_{f_i}^{\top}(t)\bsigma_{f_i}(\bx_i)
\),
one has
\(
\bfv_i(\bx_i)-\hat{\bfv}_i(\bx_i,t)
=
-\tilde{\bW}_{f_i}^{\top}(t)\bsigma_{f_i}(\bx_i)
+
\bm{\varepsilon}_{f_i}(\bx_i)
\).
Hence,
\(
\|\bfv_i(\bx_i)-\hat{\bfv}_i(\bx_i,t)\|
\leq
\bar\sigma_{f_i}\|\tilde{\bW}_{f_i}(t)\|_F
+
\bar\varepsilon_{f_i}
\).
By Theorem~\ref{lem:identification_error_i},
\(
\tilde{\bW}_{f_i}(t)
\)
is uniformly ultimately bounded. Therefore, there exists a finite
constant \(\bar W_{f_i}>0\) such that
\(
\|\tilde{\bW}_{f_i}(t)\|_F\leq\bar W_{f_i}
\),
and thus \eqref{eq:identification_error_bound_i} holds with
\(
\bar\Delta_{f_i}
:=
\bar\sigma_{f_i}\bar W_{f_i}
+
\bar\varepsilon_{f_i}
\).
\end{proof}

\section{Distributed Inverse-Optimal Fixed-Time Control}
\label{sec:inverse_optimal_control}

\begin{remark}
A central issue in fixed-time and predefined-time
optimal control is not merely to establish a
Lyapunov decay condition, but to determine how
such a condition can be systematically realized
through the optimization problem itself.
The studies in
\cite{Kokolakis2025Predefined,Gong2025,WANG2026112673}
provide important inspiration for this problem.
Nevertheless, a systematic inverse-optimal
procedure for constructing a running cost whose
minimizing feedback directly produces the
prescribed fixed-time or predefined-time decay
remains insufficiently addressed.
Indeed, imposing a desired dissipation inequality
on the optimal value function raises a fundamental
question: how should the performance index be
constructed so that its optimal solution inherently
possesses the required convergence property?

Our earlier work~\cite{VuDoan2026FixedTime}
investigated this question by reconstructing the
cost function to achieve practical fixed-time
stability under unknown nonlinear dynamics,
FDI attacks, and input saturation.
However, that construction leaves a further
challenge concerning the behavior of the
resulting feedback near the origin, particularly
the verification of the small-control property
associated with Sontag's stabilization theory
\cite{Sontag1989}.
This motivates a more systematic inverse-optimal
construction in which the prescribed Lyapunov
decay, minimum-effort feedback, and running cost
are derived consistently, with continuity at
the origin explicitly examined.

Another important limitation concerns the
generality of the system models considered.
In particular, Gong et al.~\cite{Gong2025}
address secure formation control for a structured
second-order multi-agent system.
Although useful for illustrating fixed-time RL
under FDI attacks, this setting does not establish
the same guarantees for general unknown nonlinear
leader--follower dynamics.
Such a restriction becomes important for complex
cyber-physical systems, whose nonlinear dynamics
may be difficult to characterize accurately
before controller implementation.
The present framework addresses this issue by
combining inverse-optimal feedback synthesis
with integral concurrent identification.
The desired fixed-time decay is realized through
the constructed running cost on the identified
model, while the effects of identification and
critic approximation residuals are explicitly
retained in the practical stability analysis.
\end{remark}

\subsection{Fixed-Time Optimal-Control Formulation}
Define the local augmented control vector and its local input matrix by
\begin{equation}
    \bar{\bu}_i
    :=
    \operatorname{col}
    (
    \bu_i,\bu_{\mathcal N_i}
    ),
    \qquad
    \bar{\mathbf G}_i(\boldsymbol{\xi}_i)
    :=
    \big[
        \mathbf G_{ii}(\bx_i)\;\;
        \mathbf G_{i\mathcal N}(\bx_{\mathcal N_i})
    \big].
    \label{eq:augmented_local_input}
\end{equation}

Based on \eqref{eq:local_error_compact_dynamics},
the corresponding identified local error dynamics
can be expressed compactly as
\begin{equation}
    \dot{\be}_i
    =
    \hat{\mathbf F}_i^e(\boldsymbol{\xi}_i)
    +
    \bar{\mathbf G}_i(\boldsymbol{\xi}_i)\bar{\bu}_i,
    \label{eq:identified_local_error_dynamics_compact}
\end{equation}


where
\(
\hat{\mathbf F}_i^e(\boldsymbol{\xi}_i)
:=
\sum_{j\in\mathcal N_i}
a_{ij}
\big[
\hat{\mathbf f}_i(\bx_i)
-
\hat{\mathbf f}_j(\bx_j)
-
\dot{\bh}_i
+
\dot{\bh}_j
\big]
+
b_i
\big[
\hat{\mathbf f}_i(\bx_i)
-
\mathbf f_0(\bx_0)
-
\dot{\bh}_i
\big]
\). By Lemma~\ref{lem:fixed_time}, any local feedback satisfying
\eqref{eq:prescribed_ft_decay}
renders \(\be_i=\mathbf0\) fixed-time stable with
\begin{equation}
    T_{i,\max}
    \leq
    \frac{1}{a_i(1-\mu_i)}
    +
    \frac{1}{b_i(\nu_i-1)}.
    \label{eq:local_fixed_time_bound}
\end{equation}

For follower \(i\), consider the infinite-horizon and inspired by \cite{Kokolakis2025Predefined,WANG2026112673}, the local performance index be

\begin{equation}
\begin{aligned}
J_i(\be_i,\bar{\bu}_i)
={}&
\int_0^\infty
\Big[
\mathbf L_{1i}(\be_i)\bar{\bu}_i
+
L_{2i}(\be_i)
+
\bar{\bu}_i^{\top}
\bar{\mathbf R}_i
\bar{\bu}_i
\Big]dt,
\\
&\hspace{1.5em}
\bar{\mathbf R}_i
=
\bar{\mathbf R}_i^{\top}
\succ0,
\end{aligned}
\label{eq:local_cost_function}
\end{equation}

where
\(
\mathbf L_{1i}:\Omega_i\rightarrow\mathbb R^{1\times m_i}
\)
and
\(
L_{2i}:\Omega_i\rightarrow\mathbb R
\)
are to be constructed, and
\(
m_i
\)
is the dimension of
\(
\bar{\bu}_i
\).
The term
\(\bar{\bu}_i^\top\bar{\mathbf R}_i\bar{\bu}_i\)
penalizes the local control effort,
whereas
\(\mathbf L_{1i}(\be_i)\bar{\bu}_i\)
introduces state-control coupling and
\(L_{2i}(\be_i)\) penalizes local
error-state deviations when nonnegative.
The local running cost need not be
positive definite for every
error-state-input pair. As discussed in
\cite[Sec.~8.4, pp.~524--525]{Haddad2008NonlinearControl},
the state-control cross-weighting term (\(\mathbf L_{1i}(\be_i)\bar{\bu}_i\))
may render the running cost negative
for certain control inputs without
necessarily invalidating inverse
optimality over the class of
admissible stabilizing policies.
In particular, local inverse optimality
can be established through the
corresponding HJB
equation (Section~\ref{sec3}), together with the appropriate
admissibility and transversality
conditions, even when the local running
cost is not pointwise nonnegative.
The constructed coefficients satisfy
\(\mathbf L_{1i}(\bm0)=\bm0\) and
\(L_{2i}(\bm0)=0\), as established
subsequently (See Remark~\ref{rmk3}).


Define
\(
V_i^*(\be_i)
:=
\inf_{\bar{\bu}_i}
J_i(\be_i,\bar{\bu}_i)
\),
and assume that
\(
V_i^*\in C^1(\Omega_i)
\).
For the identified local model
\eqref{eq:identified_local_error_dynamics_compact},
the corresponding Hamiltonian is

\begin{equation}
\begin{aligned}
    H_i(\be_i,\bar{\bu}_i,\nabla V_i^*)
    :={}&
    \mathbf L_{1i}(\be_i)\bar{\bu}_i
    +
    L_{2i}(\be_i)
    +
    \bar{\bu}_i^{\top}\bar{\mathbf R}_i\bar{\bu}_i
    \\
    &+
    \nabla V_i^{*\,\top}(\be_i)
    \Big[
        \hat{\mathbf F}_i^e
        +
        \bar{\mathbf G}_i\bar{\bu}_i
    \Big].
\end{aligned}
\label{eq:local_hamiltonian}
\end{equation}



Equivalently,
\[
\begin{aligned}
H_i
={}&
L_{2i}
+
\nabla V_i^{*\,\top}\hat{\mathbf F}_i^e
+
\bar{\bu}_i^{\top}\bar{\mathbf R}_i\bar{\bu}_i +
\left[
\mathbf L_{1i}
+
\nabla V_i^{*\,\top}\bar{\mathbf G}_i
\right]\bar{\bu}_i.
\end{aligned}
\]
Hence, the local HJB equation is \(\min_{\bar{\bu}_i}
H_i(\be_i,\bar{\bu}_i,\nabla V_i^*)=0.\)

Hence,
\begin{equation}
\begin{aligned}
\bar{\bu}_i^*(\be_i)
={}&
-\frac{1}{2}
\bar{\mathbf R}_i^{-1}
\Big[
\mathbf L_{1i}^{\top}(\be_i)
+
\bar{\mathbf G}_i^{\top}
\nabla V_i^*(\be_i)
\Big].
\end{aligned}
\label{eq:local_optimal_policy_general}
\end{equation}


\subsection{Reverse Construction of the Fixed-Time Optimal Cost}
\label{sec3}

For each follower $i\in\mathcal V$, let
$V_i^*:\Omega_i\to\mathbb R_{\geq 0}$
be a continuously differentiable, positive-definite
candidate Lyapunov function satisfying
$V_i^*(\mathbf{0})=0$.
Let $a_i,b_i>0$ and $\frac{1}2{}<\mu_i<1<\nu_i$
be fixed-time design parameters.
The prescribed fixed-time decay is
\begin{equation}\label{eq:prescribed_ft_decay}
\dot V_i^*(\mathbf e_i)
=
-a_i\big(V_i^*(\mathbf e_i)\big)^{\mu_i}
-b_i\big(V_i^*(\mathbf e_i)\big)^{\nu_i}.
\end{equation}

Since this is an inverse design problem, the value function
is pre-specified by the designer, and the behavior described
in~\eqref{eq:prescribed_ft_decay} is deliberately prescribed.
However, the choice of the value function is not arbitrary
but is subject to the conditions discussed in Remark~\ref{rmk3}.

\begin{assumption}[Inverse-optimal feasibility]
\label{ass:local_inverse_optimal_feasibility}
For every
\(
\mathbf e_i\in\Omega_{e_i}\setminus\{\mathbf 0\}
\),
one has
\(
D_{d,i}(\mathbf e_i)>0
\),
where
\[
\begin{aligned}
D_{d,i}(\mathbf e_i)
:={}&
\nabla V_i^{*\top}
\bar{\mathbf G}_i
\bar{\mathbf R}_i^{-1}
\bar{\mathbf G}_i^\top
\nabla V_i^*.
\end{aligned}
\]
\end{assumption}


\begin{theorem}[Local Fixed-Time Inverse-Optimal Construction]
\label{theorem2}
Consider the identified local error dynamics
\label{thm2}
\eqref{eq:identified_local_error_dynamics_compact}
and the prescribed fixed-time decay
\eqref{eq:prescribed_ft_decay}.
Suppose that
Assumption~\ref{ass:local_inverse_optimal_feasibility}
holds. Define
\(
\alpha_{d,i}(\mathbf e_i)
:=
\nabla V_i^{*\top}
\hat{\mathbf F}_i^e
+
a_i\big(V_i^*\big)^{\mu_i}
+
b_i\big(V_i^*\big)^{\nu_i}
\).
Then the unique
feedback policy
satisfying
\eqref{eq:prescribed_ft_decay}
is
\(
\bar{\mathbf u}_{c,i}(\mathbf e_i)
=
-
\frac{\alpha_{d,i}(\mathbf e_i)}
{D_{d,i}(\mathbf e_i)}
\bar{\mathbf R}_i^{-1}
\bar{\mathbf G}_i^\top
\nabla V_i^*
\).

Moreover, select the local running-cost
components as
\(
\mathbf L_{1i}^{\top}(\mathbf e_i)
=
\left[
\frac{2\alpha_{d,i}(\mathbf e_i)}
{D_{d,i}(\mathbf e_i)}
-1
\right]
\bar{\mathbf G}_i^\top
\nabla V_i^*
\)
and
\(
L_{2i}(\mathbf e_i)
=
-\nabla V_i^{*\top}
\hat{\mathbf F}_i^e
+
\frac{\alpha_{d,i}^2(\mathbf e_i)}
{D_{d,i}(\mathbf e_i)}
\).
Then the nominal local closed-loop
dynamics satisfy
\eqref{eq:prescribed_ft_decay}.
\end{theorem}

\begin{proof}
Along the identified local dynamics
\eqref{eq:identified_local_error_dynamics_compact},
the derivative of \(V_i^*\) is
\(
\dot V_i^*
=
\nabla V_i^{*\,\top}
\hat{\mathbf F}_i^e
+
\nabla V_i^{*\,\top}
\bar{\mathbf G}_i
\bar{\bu}_i
\).
Imposing the prescribed fixed-time evolution
\eqref{eq:prescribed_ft_decay}
gives
\begin{equation}
\begin{aligned}
&
\nabla V_i^{*\,\top}
\hat{\mathbf F}_i^e
+
\nabla V_i^{*\,\top}
\bar{\mathbf G}_i
\bar{\bu}_i
=
-a_i\big(V_i^*\big)^{\mu_i}
-b_i\big(V_i^*\big)^{\nu_i}.
\end{aligned}
\label{eq:local_prescribed_decay_constraint}
\end{equation}
The prescribed decay constraint
\eqref{eq:local_prescribed_decay_constraint}
is equivalently written as
\(
\nabla V_i^{*\,\top}
\bar{\mathbf G}_i
\bar{\bu}_i
=
-\alpha_{d,i}
\). Since infinitely many local augmented
controls may satisfy
\(
\nabla V_i^{*\,\top}
\bar{\mathbf G}_i
\bar{\bu}_i
+
\alpha_{d,i}
=
0
\).
Inspired by the minimum-effort stabilization
problem of Petersen and Barmish
~\cite{Petersen1987MinimumEffort}
and the pointwise min-norm CLF approach
presented in Section III of Primbs and
Giannelli~\cite{Primbs1999CLFMinNorm}.
We formulate a weighted minimum-control-effort
problem subject to the prescribed fixed-time
Lyapunov decay constraint. We select the one minimizing the
weighted local control effort
\(
\bar{\bu}_i^{\top}
\bar{\mathbf R}_i
\bar{\bu}_i
\).
Accordingly, consider
\(
\min_{\bar{\bu}_i}
\bar{\bu}_i^{\top}
\bar{\mathbf R}_i
\bar{\bu}_i
\)
subject to
\(
\nabla V_i^{*\,\top}
\bar{\mathbf G}_i
\bar{\bu}_i
+
\alpha_{d,i}
=
0
\).
The corresponding Lagrangian is
\(
\mathcal L_{c,i}
=
\bar{\bu}_i^{\top}
\bar{\mathbf R}_i
\bar{\bu}_i
+
\lambda_i
[
\nabla V_i^{*\,\top}
\bar{\mathbf G}_i
\bar{\bu}_i
+
\alpha_{d,i}
]
\).
The stationarity condition with respect
to \(\bar{\bu}_i\) gives
\(
\partial\mathcal L_{c,i}/\partial\bar{\bu}_i
=
2\bar{\mathbf R}_i\bar{\bu}_i
+
\lambda_i
\bar{\mathbf G}_i^{\top}
\nabla V_i^*
=
\mathbf0
\),
and therefore
\(
\bar{\bu}_i
=
-(\lambda_i/2)
\bar{\mathbf R}_i^{-1}
\bar{\mathbf G}_i^{\top}
\nabla V_i^*
\).
Substituting this expression into the
constraint yields
\(
-(\lambda_i/2)D_{d,i}
+
\alpha_{d,i}
=
0
\),
so that
\(
\lambda_i
=
2\alpha_{d,i}/D_{d,i}
\).
Hence, the unique local feedback policy
is given by the expression
\(\bar{\mathbf u}_{c,i}(\mathbf e_i)\)
in Theorem~\ref{thm2}.
The uniqueness follows from
\(
\bar{\mathbf R}_i\succ\mathbf0
\)
and
\(
D_{d,i}>0
\). Indeed, under the proposed feedback,
one obtains
\(
\nabla V_i^{*\,\top}
\bar{\mathbf G}_i
\bar{\bu}_{c,i}
=
-(\alpha_{d,i}/D_{d,i})
\nabla V_i^{*\,\top}
\bar{\mathbf G}_i
\bar{\mathbf R}_i^{-1}
\bar{\mathbf G}_i^{\top}
\nabla V_i^*
=
-\alpha_{d,i}
\),
and therefore
\(
\dot V_i^*
=
\nabla V_i^{*\,\top}
\hat{\mathbf F}_i^e
+
\nabla V_i^{*\,\top}
\bar{\mathbf G}_i
\bar{\bu}_{c,i}
=
\nabla V_i^{*\,\top}
\hat{\mathbf F}_i^e
-
\alpha_{d,i}
=
-a_i(V_i^*)^{\mu_i}
-b_i(V_i^*)^{\nu_i}
\). We now construct
\(
\mathbf L_{1i}(\be_i)
\)
such that the Hamiltonian-minimizing
policy \(\bar{\bu}_i^*\) coincides with
\(\bar{\bu}_{c,i}\).
From
\eqref{eq:local_optimal_policy_general},
\(
\bar{\bu}_i^*
=
-(1/2)
\bar{\mathbf R}_i^{-1}
[
\mathbf L_{1i}^{\top}
+
\bar{\mathbf G}_i^{\top}
\nabla V_i^*
]
\),
whereas
\(
\bar{\bu}_{c,i}
=
-(\alpha_{d,i}/D_{d,i})
\bar{\mathbf R}_i^{-1}
\bar{\mathbf G}_i^{\top}
\nabla V_i^*
\).
Imposing
\(
\bar{\bu}_i^*
=
\bar{\bu}_{c,i}
\)
gives
\(
\mathbf L_{1i}^{\top}
+
\bar{\mathbf G}_i^{\top}
\nabla V_i^*
=
(2\alpha_{d,i}/D_{d,i})
\bar{\mathbf G}_i^{\top}
\nabla V_i^*
\),
so that
\(
\mathbf L_{1i}^{\top}
=
[
2\alpha_{d,i}/D_{d,i}
-
1
]
\bar{\mathbf G}_i^{\top}
\nabla V_i^*
\).
This establishes the expression
for \(\mathbf L_{1i}^{\top}\).
Substituting the proposed running-cost
components into the Hamiltonian gives
the required identity. It remains to determine
\(
L_{2i}(\be_i)
\).
Substitution of
\eqref{eq:local_optimal_policy_general}
into the Hamiltonian gives
\(
H_i^*
=
L_{2i}
+
\nabla V_i^{*\,\top}
\hat{\mathbf F}_i^e
-
\frac14
[
\mathbf L_{1i}
+
\nabla V_i^{*\,\top}
\bar{\mathbf G}_i
]
\bar{\mathbf R}_i^{-1}
[
\mathbf L_{1i}^{\top}
+
\bar{\mathbf G}_i^{\top}
\nabla V_i^*
]
\).
Using
\(
\mathbf L_{1i}^{\top}
+
\bar{\mathbf G}_i^{\top}
\nabla V_i^*
=
(2\alpha_{d,i}/D_{d,i})
\bar{\mathbf G}_i^{\top}
\nabla V_i^*
\),
the quadratic term becomes
\[
\frac14
\big[
\mathbf L_{1i}
+
\nabla V_i^{*\,\top}
\bar{\mathbf G}_i
\big]
\bar{\mathbf R}_i^{-1}
\big[
\mathbf L_{1i}^{\top}
+
\bar{\mathbf G}_i^{\top}
\nabla V_i^*
\big]
=
\frac{
\alpha_{d,i}^2(\be_i)
}{
D_{d,i}(\be_i)
}.
\]

Hence,
\(
H_i^*
=
L_{2i}
+
\nabla V_i^{*\,\top}
\hat{\mathbf F}_i^e
-
\alpha_{d,i}^2/D_{d,i}
\).
The HJB condition
\(
H_i^*=0
\)
therefore gives
\(
L_{2i}
=
-
\nabla V_i^{*\,\top}
\hat{\mathbf F}_i^e
+
\alpha_{d,i}^2/D_{d,i}
\),
which establishes the compact expression
for the local running-cost component
\(L_{2i}\).
Expanding
\(
\alpha_{d,i}
\)
and
\(
D_{d,i}
\)
yields the equivalent expanded expression
for \(L_{2i}\), completing the proof.
\end{proof}

\begin{remark}
    Theorem \ref{theorem2} clarifies the process of constructing the cost function, which was not explicitly addressed in \cite{Kokolakis2025Predefined,WANG2026112673}.
\end{remark}

\begin{remark}[Small-control property at the origin~\cite{Vu2026PredefinedIRL}]
\label{rmk3}
Since
\(
\bar{\mathbf R}_i\succ0
\),
the condition
\(
D_{d,i}(\mathbf e_i)=0
\)
holds if and only if
\(
\bar{\mathbf G}_i^\top
\nabla V_i^*=\mathbf 0
\).
This occurs in two cases:
(i) \(\nabla V_i^*=\mathbf 0\),
or (ii)
\(
\nabla V_i^*\neq\mathbf 0
\)
but belongs to the null space of
\(
\bar{\mathbf G}_i^\top
\).
Assumption~\ref{ass:local_inverse_optimal_feasibility}
excludes both cases away from the origin.
Following the small-control property of
Sontag~\cite{Sontag1989}, continuity of the
feedback at the origin can be established
under suitable local regularity conditions.
In the present inverse-optimal construction,
the candidate value function is freely designed.
For example, let
\(
V_i^*
=
\frac12\mathbf e_i^\top
\mathbf P_i\mathbf e_i
\),
where
\(
\mathbf P_i\succ0
\).
If the identified nominal drift is locally
Lipschitz and vanishes at the origin,
\(
D_{d,i}(\mathbf e_i)
\geq c_i\|\mathbf e_i\|^2
\)
for some \(c_i>0\), and the input matrices
are locally bounded, then
\(
\|\bar{\mathbf u}_{c,i}(\mathbf e_i)\|
=
O(\|\mathbf e_i\|^{2\mu_i-1})
\).
Since \(1/2<\mu_i<1\), it follows that
\(
\lim_{\mathbf e_i\to\mathbf 0}
\bar{\mathbf u}_{c,i}(\mathbf e_i)=\mathbf 0
\).
Thus,
\(
\bar{\mathbf u}_{c,i}(\mathbf 0)=\mathbf 0
\)
defines a continuous extension at the origin.
Moreover, under the preceding regularity conditions,
the corresponding inverse-optimal running-cost
components also admit continuous extensions,
satisfying
\(
\lim_{\mathbf e_i\to\mathbf 0}
\mathbf L_{1i}(\mathbf e_i)=\mathbf 0
\)
and
\(
\lim_{\mathbf e_i\to\mathbf 0}
L_{2i}(\mathbf e_i)=0
\).
Consequently,
\(
\mathbf L_{1i}(\mathbf 0)=\mathbf 0
\)
and
\(
L_{2i}(\mathbf 0)=0
\)
complete the continuous extension of the
inverse-optimal feedback and its associated
running-cost components at the origin.
\end{remark}

\subsection{Distributed Critic-Only Integral Reinforcement Learning}
\label{subsec:distributed_critic_irl}

\begin{remark}
Although the unknown follower drift
\(\mathbf f_i(\bx_i)\) is reconstructed in the
preceding identification stage, its estimate
\(\hat{\mathbf f}_i(\bx_i)\) is not treated as
an exact plant model.
Instead, the identified local drift
\(\hat{\mathbf F}_i^e\) is used to construct
the inverse-optimal running-cost components
\(\mathbf L_{1i}\) and \(L_{2i}\) and prescribe
the nominal fixed-time decay of
\(V_i^*(\be_i)\).
Similarly, Wang et al.~\cite{Wang2024FixedTimeFormation}
employed neural identification to support
RL-based fixed-time formation control.
In the subsequent IRL formulation,
the actual drift \(\mathbf F_i^e\) remains
unknown, while the critic weights are updated
using integral Bellman errors constructed from
measured state and input trajectories.
Specifically, the integral formulation employs
finite-window differences of the critic basis
functions and integrals of known quantities.
The identification discrepancy
\(\mathbf F_i^e-\hat{\mathbf F}_i^e\),
together with the critic approximation error,
is explicitly accounted for in the residual
terms of the practical fixed-time convergence
analysis.
\end{remark}

After constructing the inverse-optimal functions
\(
\mathbf L_{1i}(\be_i)
\)
and
\(
L_{2i}(\be_i)
\), consider the true local synchronization-error dynamics
\(
\dot{\be}_i
=
\mathbf F_i^{e}(\boldsymbol{\xi}_i)
+
\bar{\mathbf G}_i(\boldsymbol{\xi}_i)\bar{\bu}_i
\). Similar to Eq.~\eqref{eq:local_optimal_policy_general}, the optimal control policy for the case where the model is the true model is

\begin{equation}
\begin{aligned}
\bar{\bu}_i^{*}(\be_i)
=
-\frac{1}{2}\bar{\bR}_i^{-1}
\big[
\mathbf L_{1i}(\be_i)
+
\bar{\mathbf G}_i^{T}
\nabla V_i^{*}(\be_i)
\big].
\end{aligned}
\label{eq:local_optimal_policy}
\end{equation}

Note that the optimal policy in Eq.~\eqref{eq:local_optimal_policy} is the optimal policy based on the true model, it differs from the optimal policy in Eq~\eqref{eq:local_optimal_policy_general}.

Motivated by the Weierstrass approximation theorem~\cite{Vamvoudakis2010}
and under suitable smooth approximation assumptions,
on the compact set
\(
\Omega_{e_i}
\),
let
\begin{equation}
V_i^{*}(\be_i)
=
\bW_i^{*T}
\bphi_i(\be_i)
+
\varepsilon_i(\be_i),
\label{eq:value_approximation}
\end{equation}

where
\(
\bphi_i(\be_i)\in\mathbb R^{N_{c_i}}
\),
\(
\bW_i^{*}\in\mathbb R^{N_{c_i}}
\),
and
\(
\varepsilon_i(\be_i)
\)
denotes the value-function approximation error.
Therefore,
\(
\nabla V_i^{*}(\be_i)
=
(
\nabla\bphi_i(\be_i)
)^{T}
\bW_i^{*}
+
\nabla\varepsilon_i(\be_i)
\).
The critic estimate is
\begin{equation}
\hat V_i(\be_i,\hat{\bW_i})
=
\hat{\bW}_i^{T}
\bphi_i(\be_i),
\qquad
\nabla\hat V_i(\be_i,\hat{\bW_i})
=
\big(
\nabla\bphi_i(\be_i)
\big)^{T}
\hat{\bW}_i .
\label{eq:critic_estimate}
\end{equation}

Thus, the critic-generated policy is
\begin{equation}
\hat{\bar{\bu}}_i(\be_i,\hat{\bW_i})
=
-\frac12
\bar{\bR}_i^{-1}
\left[
\mathbf L_{1i}^{T}(\be_i)
+
\bar{\mathbf G}_i^{T}
\big(
\nabla\bphi_i(\be_i)
\big)^{T}
\hat{\bW}_i
\right].
\label{eq:critic_generated_policy}
\end{equation}


\begin{proposition}[Critic approximation residual~\cite{Vamvoudakis2010,Kokolakis2025Predefined}]
\label{prop:critic_approximation_residual}
On the compact set
\(
\Omega_{e_i}
\),
suppose that the critic approximation converges
uniformly in the
\(
C^1
\)
sense as
\(
N_{C_i}\to\infty
\).
Then,
\[
\lim_{N_{C_i}\to\infty}
\sup_{\be_i\in\Omega_{e_i}}
\left(
|\varepsilon_i(\be_i)|
+
\|\nabla\varepsilon_i(\be_i)\|
\right)
=0.
\]
\end{proposition}


\begin{remark}[Approximation accuracy and computational trade-off]
Proposition~\ref{prop:critic_approximation_residual}
provides one possible means of reducing the
ultimate bounds established in Theorems~\ref{thm:critic_fixed_time} and~\ref{thm:network_fixed_time}
by increasing the number of critic neurons
\(
N_{C_i}
\).
However, approximation residuals are only one
of several factors determining the UUB radii.
Increasing
\(
N_{C_i}
\)
may improve approximation accuracy at the
expense of computational complexity, whereas
a smaller network may sacrifice approximation
capability~\cite{ParkSandberg1991}.
This trade-off suggests a possible extension
using an adaptive network architecture, in which
\(
N_{C_i}
\)
is adjusted online through neuron allocation
or growing-and-pruning mechanisms, as explored
in~\cite{Platt1991RAN,Han2010SORBF}.
\end{remark}

During learning, a bounded probing signal
\(
\bar{\mathbf e}_{u_i}(t)
\)
is introduced to provide sufficient excitation
for critic-weight adaptation.
The behavior input is given by
\(
\bar{\bu}_i(t)
=
\hat{\bar{\bu}}_i(t)
+
\bar{\mathbf e}_{u_i}(t)
\).
Thus,
\(
\bar{\bu}_i-\bar{\bu}_i^*
=
(\bar{\bu}_i-\hat{\bar{\bu}}_i)
+
(\hat{\bar{\bu}}_i-\bar{\bu}_i^*)
=
\bar{\mathbf e}_{u_i}
+
(\hat{\bar{\bu}}_i-\bar{\bu}_i^*)
\).


Along the measured local dynamics
\(
\dot{\be}_i
=
\mathbf F_i^e(\boldsymbol{\xi}_i)
+
\bar{\mathbf G}_i(\boldsymbol{\xi}_i)
\bar{\bu}_i
\),
the optimal HJB equation associated with
the actual local dynamics is
\(
0
=
\ell_i(\be_i,\bar{\bu}_i^*)
+
\nabla V_i^{*T}(\be_i)
[
\mathbf F_i^e(\boldsymbol{\xi}_i)
+
\bar{\mathbf G}_i(\boldsymbol{\xi}_i)
\bar{\bu}_i^*
]
\).
Define the local policy-suboptimality residual
\(
\delta_{\hat u_i}(\be_i,t)
:=
\ell_i(\be_i,\hat{\bar{\bu}}_i)
-
\ell_i(\be_i,\bar{\bu}_i^*)
+
\nabla V_i^{*T}(\be_i)
\bar{\mathbf G}_i(\boldsymbol{\xi}_i)
(\hat{\bar{\bu}}_i-\bar{\bu}_i^*)
\).
Consequently,
\(
\nabla V_i^{*T}
[
\mathbf F_i^e
+
\bar{\mathbf G}_i\hat{\bar{\bu}}_i
]
=
-\ell_i(\be_i,\hat{\bar{\bu}}_i)
+
\delta_{\hat u_i}(\be_i,t)
\).
Using
\(
\mathbf F_i^e
+
\bar{\mathbf G}_i\bar{\bu}_i
=
\mathbf F_i^e
+
\bar{\mathbf G}_i\hat{\bar{\bu}}_i
+
\bar{\mathbf G}_i
(\bar{\bu}_i-\hat{\bar{\bu}}_i)
\),
one obtains
\(
\dot V_i^*
=
-\ell_i(\be_i,\hat{\bar{\bu}}_i)
+
\delta_{\hat u_i}(\be_i,t)
+
\nabla V_i^{*T}
\bar{\mathbf G}_i
(\bar{\bu}_i-\hat{\bar{\bu}}_i)
\).
Integrating over
\(
[t-T_i,t]
\)
gives
\begin{equation}
\begin{aligned}
0
={}&
V_i^*(\be_i(t))
-
V_i^*(\be_i(t-T_i))
+
\int_{t-T_i}^{t}
\ell_i
\big(
\be_i(\tau),
\hat{\bar{\bu}}_i(\tau)
\big)d\tau
\\
&-
\int_{t-T_i}^{t}
\nabla V_i^{*T}(\be_i(\tau))
\bar{\mathbf G}_i(\boldsymbol{\xi}_i(\tau))
\big[
\bar{\bu}_i(\tau)
-
\hat{\bar{\bu}}_i(\tau)
\big]d\tau
\\
&-
\int_{t-T_i}^{t}
\delta_{\hat u_i}(\be_i(\tau),\tau)
\,d\tau .
\end{aligned}
\label{eq:integral_bellman}
\end{equation}

From
\eqref{eq:value_approximation},
\begin{equation}
\begin{aligned}
&
V_i^*(\be_i(t))
-
V_i^*(\be_i(t-T_i))
\\
&\qquad=
\bW_i^{*T}
\Delta\bphi_i(t)
+
\varepsilon_i(\be_i(t))
-
\varepsilon_i(\be_i(t-T_i)),
\end{aligned}
\label{eq:value_difference}
\end{equation}
where
\begin{equation}
\Delta\bphi_i(t)
:=
\bphi_i(\be_i(t))
-
\bphi_i(\be_i(t-T_i)).
\label{eq:delta_phi_i}
\end{equation}

Substituting
\(
\nabla V_i^*
=
(\nabla\bphi_i)^T\bW_i^*
+
\nabla\varepsilon_i
\)
into
\eqref{eq:integral_bellman}
gives
\begin{equation}
\bW_i^{*T}
\bOmega_i(t)
+
\rho_i(t)
+
\varepsilon_{B_i}(t)
=
0,
\label{eq:integral_bellman_regression}
\end{equation}
where
\begin{equation}
\begin{aligned}
\bOmega_i(t)
:={}&
\Delta\bphi_i(t)
-
\int_{t-T_i}^{t}
\nabla\bphi_i(\be_i(\tau))
\bar{\mathbf G}_i(\boldsymbol{\xi}_i(\tau))
\big[
\bar{\bu}_i(\tau)
-
\hat{\bar{\bu}}_i(\tau)
\big]d\tau ,
\end{aligned}
\label{eq:Omega_i}
\end{equation}
and
\begin{equation}
\rho_i(t)
:=
\int_{t-T_i}^{t}
\ell_i
\big(
\be_i(\tau),
\hat{\bar{\bu}}_i(\tau)
\big)d\tau .
\label{eq:rho_i}
\end{equation}

The aggregate local Bellman residual is
\begin{equation}
\begin{aligned}
\varepsilon_{B_i}(t)
:={}&
\varepsilon_i(\be_i(t))
-
\varepsilon_i(\be_i(t-T_i))
-
\int_{t-T_i}^{t}
\nabla\varepsilon_i^T(\be_i(\tau))
\bar{\mathbf G}_i(\boldsymbol{\xi}_i(\tau))
\\
&\qquad\times
\big[
\bar{\bu}_i(\tau)
-
\hat{\bar{\bu}}_i(\tau)
\big]d\tau
-
\int_{t-T_i}^{t}
\delta_{\hat u_i}(\be_i(\tau),\tau)
\,d\tau .
\end{aligned}
\label{eq:bellman_approximation_residual}
\end{equation}
Hence,
\(
\varepsilon_{B_i}(t)
\)
contains the value-function approximation
error and the local policy-suboptimality
residual, without an explicit identification
discrepancy term.

For the critic estimate
\(
\hat V_i
\),
define the integral Bellman residual
\begin{equation}
\xi_i(t)
:=
\hat{\bW}_i^{T}
\bOmega_i(t)
+
\rho_i(t).
\label{eq:integral_Bellman_residual}
\end{equation}

Let
\(
\tilde{\bW}_i
:=
\bW_i^{*}
-
\hat{\bW}_i
\).
Under exact approximation,
\(
\varepsilon_{B_i}=0
\),
and hence
\begin{equation}
\xi_i(t)
=
-\tilde{\bW}_i^{T}
\bOmega_i(t).
\label{eq:bellman_residual_exact}
\end{equation}

More generally, when
\(
\varepsilon_{B_i}\neq0
\),
\begin{equation}
\xi_i(t)
=
-\tilde{\bW}_i^{T}
\bOmega_i(t)
-
\varepsilon_{B_i}(t).
\label{eq:bellman_residual_general}
\end{equation}

To normalize the regressor, define
\begin{equation}
\mathscr M_i(t)
:=
1+
\bOmega_i^{T}(t)
\bOmega_i(t),
\qquad
\bar{\bOmega}_i(t)
:=
\frac{\bOmega_i(t)}
{\mathscr M_i(t)} .
\label{eq:normalization}
\end{equation}

The normalized Bellman residual is
\begin{equation}
\begin{aligned}
\mathscr B_i(t)
:={}&
\frac{\xi_i(t)}
{\mathscr M_i(t)}
={}
\hat{\bW}_i^{T}
\bar{\bOmega}_i(t)
+
\frac{\rho_i(t)}
{\mathscr M_i(t)}.
\end{aligned}
\label{eq:normalized_Bellman_error}
\end{equation}

Under exact approximation,
\(
\mathscr B_i(t)
=
-\tilde{\bW}_i^{T}
\bar{\bOmega}_i(t)
\).
In the general approximation case, define
\(
\bar\varepsilon_{B_i}(t)
:=
\varepsilon_{B_i}(t)/\mathscr M_i(t)
\).
Then
\begin{equation}
\mathscr B_i(t)
=
-\tilde{\bW}_i^{T}
\bar{\bOmega}_i(t)
-
\bar\varepsilon_{B_i}(t).
\label{eq:normalized_Bellman_error_general}
\end{equation}

Let
\(
M_i
\)
previously recorded integral samples be denoted by
\(
\{
\bOmega_{i,j},
\rho_{i,j}
\}_{j=1}^{M_i}
\).
For each stored sample, define
\(
\mathscr M_{i,j}
:=
1+
\bOmega_{i,j}^{T}
\bOmega_{i,j}
\)
and
\(
\bar{\bOmega}_{i,j}
:=
\bOmega_{i,j}/\mathscr M_{i,j}
\).

The corresponding normalized stored Bellman residual is
\begin{equation}
\mathscr B_{i,j}
:=
\frac{
\hat{\bW}_i^{T}\bOmega_{i,j}
+
\rho_{i,j}
}{
\mathscr M_{i,j}
}
=
\hat{\bW}_i^{T}\bar{\bOmega}_{i,j}
+
\frac{\rho_{i,j}}{\mathscr M_{i,j}}.
\label{eq:stored_error}
\end{equation}



For each recorded sample, define
\(
\bar\varepsilon_{B_i,j}
:=
\varepsilon_{B_i,j}/\mathscr M_{i,j}
\).
Then
\(
\mathscr B_{i,j}
=
-\tilde{\bW}_i^{T}
\bar{\bOmega}_{i,j}
-
\bar\varepsilon_{B_i,j}
\).
Assume that the approximation residual is bounded on the compact operating set so that there exists
\(
\mathscr K_{B_i}>0
\)
satisfying
\(
|\bar\varepsilon_{B_i}(t)|
\leq
\mathscr K_{B_i}
\)
and
\(
|\bar\varepsilon_{B_i,j}|
\leq
\mathscr K_{B_i}
\)
for
\(
j=1,\ldots,M_i
\).

The recorded regressors are assumed to satisfy the following finite informativity condition.

\begin{assumption}[Finite replay informativity~\cite{Modares2014,Kokolakis2025Predefined}]
\label{ass:replay_ie}
For each follower
\(
i
\),
there exists a constant
\(
\alpha_i>0
\)
such that
\begin{equation}
\sum_{j=1}^{M_i}
\bar{\bOmega}_{i,j}
\bar{\bOmega}_{i,j}^{T}
\succeq
\alpha_i
I_{N_{c_i}}.
\label{eq:replay_ie}
\end{equation}
Equivalently,
\(
\lambda_{\min}
(
\sum_{j=1}^{M_i}
\bar{\bOmega}_{i,j}
\bar{\bOmega}_{i,j}^{T}
)
\geq
\alpha_i
>0
\).
\end{assumption}

Let
\(
\frac12<\mu_i<1<\nu_i
\),
\(
\gamma_1=2\mu_i-1
\),
\(
\gamma_2=2\nu_i-1
\),
and define
\(
\operatorname{sig}^{\gamma}(s)
:=
|s|^{\gamma}
\operatorname{sgn}(s)
\).
Consider the scalar two-power penalty
\(
\Psi_i(s)
:=
\frac{k_{i1}}{\gamma_1+1}
|s|^{\gamma_1+1}
+
\frac{k_{i2}}{\gamma_2+1}
|s|^{\gamma_2+1}
\),
where
\(
k_{i1},k_{i2}>0
\).
Its derivative satisfies
\(
\partial\Psi_i(s)/\partial s
=
k_{i1}
\operatorname{sig}^{\gamma_1}(s)
+
k_{i2}
\operatorname{sig}^{\gamma_2}(s)
\).

Since
\(
\mathscr B_i(t)
=
\hat{\bW}_i^{T}
\bar{\bOmega}_i(t)
+
\rho_i(t)/\mathscr M_i(t)
\),
one has
\(
\partial\mathscr B_i(t)/
\partial\hat{\bW}_i
=
\bar{\bOmega}_i(t)
\).
Therefore,
\(
\nabla_{\hat{\bW}_i}
\Psi_i(\mathscr B_i)
=
\bar{\bOmega}_i
[
k_{i1}
\operatorname{sig}^{\gamma_1}(\mathscr B_i)
+
k_{i2}
\operatorname{sig}^{\gamma_2}(\mathscr B_i)
]
\).
Similarly,
\(
\nabla_{\hat{\bW}_i}
\Psi_i(\mathscr B_{i,j})
=
\bar{\bOmega}_{i,j}
[
k_{i1}
\operatorname{sig}^{\gamma_1}(\mathscr B_{i,j})
+
k_{i2}
\operatorname{sig}^{\gamma_2}(\mathscr B_{i,j})
]
\).

Define the composite critic objective
\begin{equation}
J_{W_i}
:=
\Psi_i(\mathscr B_i)
+
\sum_{j=1}^{M_i}
\Psi_i(\mathscr B_{i,j}).
\label{eq:critic_objective}
\end{equation}

Then
\(
\nabla_{\hat{\bW}_i}
J_{W_i}
=
\bar{\bOmega}_i
[
k_{i1}
\operatorname{sig}^{\gamma_1}(\mathscr B_i)
+
k_{i2}
\operatorname{sig}^{\gamma_2}(\mathscr B_i)
]
+
\sum_{j=1}^{M_i}
\bar{\bOmega}_{i,j}
[
k_{i1}
\operatorname{sig}^{\gamma_1}(\mathscr B_{i,j})
+
k_{i2}
\operatorname{sig}^{\gamma_2}(\mathscr B_{i,j})
]
\).

Using the gradient flow
\(
\dot{\hat{\bW}}_i
=
-\bGamma_i
\nabla_{\hat{\bW}_i}
J_{W_i}
\),
with
\(
\bGamma_i
=
\bGamma_i^{T}
\succ0
\),
gives the critic update law
\begin{equation}
\begin{aligned}
\dot{\hat{\bW}}_i
=
-\bGamma_i
\Bigg\{
&
\bar{\bOmega}_i
\left[
k_{i1}
\operatorname{sig}^{\gamma_1}(\mathscr B_i)
+
k_{i2}
\operatorname{sig}^{\gamma_2}(\mathscr B_i)
\right]
\\
&+
\sum_{j=1}^{M_i}
\bar{\bOmega}_{i,j}
\left[
k_{i1}
\operatorname{sig}^{\gamma_1}(\mathscr B_{i,j})
+
k_{i2}
\operatorname{sig}^{\gamma_2}(\mathscr B_{i,j})
\right]
\Bigg\}.
\end{aligned}
\label{eq:critic_update}
\end{equation}

\subsection{Fixed-Time Integral Critic Learning}
\label{subsec:fixed_time_critic}

Define
\(
V_{W_i}
:=
\frac12
\tilde{\bW}_i^{T}
\bGamma_i^{-1}
\tilde{\bW}_i
\) and recall that \(
\frac12<\mu_i<1<\nu_i
\),
\(
\gamma_1=2\mu_i-1
\),
\(
\gamma_2=2\nu_i-1
\).

\begin{theorem}[Fixed-time convergence of the distributed integral critic]
\label{thm:critic_fixed_time}

Suppose that
Assumption~\ref{ass:replay_ie}
holds and that the critic weights are updated according to
\eqref{eq:critic_update}.
Define
\begin{equation}
\begin{aligned}
C_{W_i1}
:={}&
\frac{k_{i1}}{2}
\alpha_i^{\mu_i}
\left[
2\lambda_{\min}(\bGamma_i)
\right]^{\mu_i},
\\
C_{W_i2}
:={}&
\frac{k_{i2}}{2}
M_i^{1-\nu_i}
\alpha_i^{\nu_i}
\left[
2\lambda_{\min}(\bGamma_i)
\right]^{\nu_i}.
\end{aligned}
\label{eq:critic_fixed_time_coefficients}
\end{equation}

Then the following statements hold.

\emph{(i) Exact-residual case:}
if
\(
\bar\varepsilon_{B_i}(t)=0
\)
and
\(
\bar\varepsilon_{B_i,j}=0
\)
for
\(
j=1,\ldots,M_i
\),
then
\begin{equation}
\tilde{\bW}_i(t)
=
\mathbf0,
\qquad
t\geq T_{W_i}^{0},
\label{eq:critic_exact_fixed_time_convergence}
\end{equation}
where
\begin{equation}
T_{W_i}^{0}
\leq
\frac{1}
{C_{W_i1}(1-\mu_i)}
+
\frac{1}
{C_{W_i2}(\nu_i-1)}
\label{eq:critic_exact_fixed_time_bound}
\end{equation}

denotes the exact fixed-time convergence bound
for follower \(i\) after finite informativity
is established.


\emph{(ii) Nonzero-residual case:}
suppose that there exists a finite constant
\(
\mathscr K_{B_i}>0
\)
such that
\(
|\bar\varepsilon_{B_i}(t)|
\leq
\mathscr K_{B_i}
\)
and
\(
|\bar\varepsilon_{B_i,j}|
\leq
\mathscr K_{B_i}
\)
for
\(j=1,\ldots,M_i\).
Then there exists a finite constant
\(
\Delta_{W_i}\geq0
\),
which can be chosen as
\(
\Delta_{W_i}
=
(M_i+1)
\big[
k_{i1}d_{2\mu_i}\mathscr K_{B_i}^{2\mu_i}
+
k_{i2}d_{2\nu_i}\mathscr K_{B_i}^{2\nu_i}
\big]
\),
where
\(
d_p=L_p(L_p+1)^{p-1}+\frac12L_p^p
\)
and
\(
L_p=2\max\{1,p-1\}
\)
for
\(p>1\).
Then, for any
\(0<\chi_{W_i}<1\),
the critic Lyapunov function enters the forward-invariant set

\begin{equation}
\mathcal B_{W_i}
:=
\left\{
\tilde{\bW}_i:
V_{W_i}
\leq
\bar V_{W_i}
\right\},
\label{eq:critic_fixed_time_residual_set}
\end{equation}
where
\(
\bar V_{W_i}>0
\)
is the unique positive solution of
\begin{equation}
C_{W_i1}
\bar V_{W_i}^{\mu_i}
+
C_{W_i2}
\bar V_{W_i}^{\nu_i}
=
\frac{\Delta_{W_i}}
{1-\chi_{W_i}},
\label{eq:critic_fixed_time_residual_equation}
\end{equation}
within the uniform time
\begin{equation}
T_{W_i}
\leq
\frac{1}
{\chi_{W_i}C_{W_i1}(1-\mu_i)}
+
\frac{1}
{\chi_{W_i}C_{W_i2}(\nu_i-1)}.
\label{eq:critic_practical_fixed_time_bound}
\end{equation}
Consequently,
\begin{equation}
\|
\tilde{\bW}_i(t)
\|
\leq
\sqrt{
2\lambda_{\max}(\bGamma_i)
\bar V_{W_i}
},
\qquad
t\geq T_{W_i}.
\label{eq:critic_fixed_time_weight_bound}
\end{equation}
\end{theorem}

\begin{proof}

Consider
\(
V_{W_i}
=
\frac12
\tilde{\bW}_i^{T}
\bGamma_i^{-1}
\tilde{\bW}_i
\).
Since
\(
\bW_i^*
\)
is constant,
\(
\dot{\tilde{\bW}}_i
=
-\dot{\hat{\bW}}_i
\),
and therefore
\(
\dot V_{W_i}
=
\tilde{\bW}_i^{T}
\bGamma_i^{-1}
\dot{\tilde{\bW}}_i
\).
Substituting
\eqref{eq:critic_update}
gives
\(
\begin{aligned}
\dot V_{W_i}
={}&
\tilde{\bW}_i^{T}
\bar{\bOmega}_i
\left[
k_{i1}
\operatorname{sig}^{\gamma_1}(\mathscr B_i)
+
k_{i2}
\operatorname{sig}^{\gamma_2}(\mathscr B_i)
\right]
\\
&+
\sum_{j=1}^{M_i}
\tilde{\bW}_i^{T}
\bar{\bOmega}_{i,j}
\left[
k_{i1}
\operatorname{sig}^{\gamma_1}(\mathscr B_{i,j})
+
k_{i2}
\operatorname{sig}^{\gamma_2}(\mathscr B_{i,j})
\right].
\end{aligned}
\)

Using
\(
\mathscr B_i
=
-\tilde{\bW}_i^{T}
\bar{\bOmega}_i
-
\bar\varepsilon_{B_i}
\)
and
\(
\mathscr B_{i,j}
=
-\tilde{\bW}_i^{T}
\bar{\bOmega}_{i,j}
-
\bar\varepsilon_{B_i,j}
\),

one obtains
\begin{equation*}
\begin{aligned}
\dot V_{W_i}
={}&
-k_{i1}|\mathscr B_i|^{\gamma_1+1}
-k_{i2}|\mathscr B_i|^{\gamma_2+1}
\\
&-
k_{i1}
\sum_{j=1}^{M_i}
|\mathscr B_{i,j}|^{\gamma_1+1}
-
k_{i2}
\sum_{j=1}^{M_i}
|\mathscr B_{i,j}|^{\gamma_2+1}
\\
&-
\bar\varepsilon_{B_i}
\left[
k_{i1}
\operatorname{sig}^{\gamma_1}(\mathscr B_i)
+
k_{i2}
\operatorname{sig}^{\gamma_2}(\mathscr B_i)
\right]
\\
&-
\sum_{j=1}^{M_i}
\bar\varepsilon_{B_i,j}
\left[
k_{i1}
\operatorname{sig}^{\gamma_1}(\mathscr B_{i,j})
+
k_{i2}
\operatorname{sig}^{\gamma_2}(\mathscr B_{i,j})
\right].
\end{aligned}
\end{equation*}

In the exact-residual case,
\(
\bar\varepsilon_{B_i}=0
\)
and
\(
\bar\varepsilon_{B_i,j}=0
\),
so

\(
\mathscr B_{i,j}
=
-\tilde{\bW}_i^{T}
\bar{\bOmega}_{i,j}
\).
Discarding the nonpositive current online terms yields
\(
\dot V_{W_i}
\leq
-k_{i1}
\sum_{j=1}^{M_i}
\left|
\tilde{\bW}_i^{T}
\bar{\bOmega}_{i,j}
\right|^{\gamma_1+1}
-
k_{i2}
\sum_{j=1}^{M_i}
\left|
\tilde{\bW}_i^{T}
\bar{\bOmega}_{i,j}
\right|^{\gamma_2+1}
\).

Define
\(
z_{i,j}:=\tilde{\bW}_i^{T}\bar{\bOmega}_{i,j}
\).
By Assumption~\ref{ass:replay_ie},
\(
\sum_{j=1}^{M_i}|z_{i,j}|^2
\geq
\alpha_i\|\tilde{\bW}_i\|^2
\).
Since
\(
\gamma_1+1=2\mu_i
\),
one obtains
\(
\sum_{j=1}^{M_i}|z_{i,j}|^{\gamma_1+1}
\geq
\alpha_i^{\mu_i}
\|\tilde{\bW}_i\|^{2\mu_i}
\).
Similarly, using
\(
\gamma_2+1=2\nu_i
\)
gives
\(
\sum_{j=1}^{M_i}|z_{i,j}|^{\gamma_2+1}
\geq
M_i^{1-\nu_i}
\alpha_i^{\nu_i}
\|\tilde{\bW}_i\|^{2\nu_i}
\).

Furthermore,
\(
V_{W_i}
=
\frac12
\tilde{\bW}_i^{T}
\bGamma_i^{-1}
\tilde{\bW}_i
\leq
[2\lambda_{\min}(\bGamma_i)]^{-1}
\|\tilde{\bW}_i\|^2
\),
and hence
\(
\|\tilde{\bW}_i\|^2
\geq
2\lambda_{\min}(\bGamma_i)V_{W_i}
\).
Consequently,
\(
\|\tilde{\bW}_i\|^{2\mu_i}
\geq
[2\lambda_{\min}(\bGamma_i)]^{\mu_i}
V_{W_i}^{\mu_i}
\)
and
\(
\|\tilde{\bW}_i\|^{2\nu_i}
\geq
[2\lambda_{\min}(\bGamma_i)]^{\nu_i}
V_{W_i}^{\nu_i}
\).

Consequently,
\begin{equation}
\dot V_{W_i}
\leq
-C_{W_i1}
V_{W_i}^{\mu_i}
-C_{W_i2}
V_{W_i}^{\nu_i}.
\label{eq:critic_exact_fixed_time_decay}
\end{equation}
Since
\(
0<\mu_i<1<\nu_i
\),
Lemma~\ref{lem:fixed_time}
gives
\(
T_{W_i}^{0}
\leq
\frac{1}
{C_{W_i1}(1-\mu_i)}
+
\frac{1}
{C_{W_i2}(\nu_i-1)}
\). Hence,
\(
\tilde{\bW}_i(t)=\mathbf0
\)
for all
\(
t\geq T_E+T_{W_i}^{0}
\).

For the nonzero-residual case, the normalized Bellman approximation residuals satisfy
\(
|\bar\varepsilon_{B_i}(t)|\leq\mathscr K_{B_i}
\)
and
\(
|\bar\varepsilon_{B_i,j}|\leq\mathscr K_{B_i}
\).
Applying the signed-power inequality
\(
z\operatorname{sig}^{p-1}(-z-\varepsilon)
\leq
-\frac12|z|^p+d_p|\varepsilon|^p
\)~\cite{Gong2025},
for \(p>1\), with
\(
p=\gamma_1+1=2\mu_i
\)
and
\(
p=\gamma_2+1=2\nu_i
\),
gives
\(
\dot V_{W_i}
\leq
-\frac{k_{i1}}{2}
\sum_{j=1}^{M_i}|z_{i,j}|^{2\mu_i}
-\frac{k_{i2}}{2}
\sum_{j=1}^{M_i}|z_{i,j}|^{2\nu_i}
+\Delta_{W_i}
\).

By Assumption~\ref{ass:replay_ie} and the power-sum inequalities,
\(
\sum_{j=1}^{M_i}|z_{i,j}|^{2\mu_i}
\geq
\alpha_i^{\mu_i}
\|\tilde{\bW}_i\|^{2\mu_i}
\)
and
\(
\sum_{j=1}^{M_i}|z_{i,j}|^{2\nu_i}
\geq
M_i^{1-\nu_i}\alpha_i^{\nu_i}
\|\tilde{\bW}_i\|^{2\nu_i}
\).
Consequently,
\(
\dot V_{W_i}
\leq
-\frac{k_{i1}}{2}
\alpha_i^{\mu_i}
\|\tilde{\bW}_i\|^{2\mu_i}
-\frac{k_{i2}}{2}
M_i^{1-\nu_i}\alpha_i^{\nu_i}
\|\tilde{\bW}_i\|^{2\nu_i}
+\Delta_{W_i}
\).

Using
\(
\|\tilde{\bW}_i\|^2
\geq
2\lambda_{\min}(\bGamma_i)V_{W_i}
\)
and the definitions of
\(C_{W_i1}\) and \(C_{W_i2}\),
one obtains
\begin{equation}
\dot V_{W_i}
\leq
-C_{W_i1}V_{W_i}^{\mu_i}
-C_{W_i2}V_{W_i}^{\nu_i}
+\Delta_{W_i}.
\label{eq:critic_practical_fixed_time_derivative}
\end{equation}

For
\(
0<\chi_{W_i}<1
\),
let
\(
\bar V_{W_i}
\)
satisfy
\eqref{eq:critic_fixed_time_residual_equation}.
Since
\(
C_{W_i1},C_{W_i2}>0
\)
and
\(
\mu_i,\nu_i>0
\),
the function
\(
C_{W_i1}V_{W_i}^{\mu_i}
+
C_{W_i2}V_{W_i}^{\nu_i}
\)
is strictly increasing for
\(
V_{W_i}\geq0
\).
Thus, whenever
\(
V_{W_i}>\bar V_{W_i}
\),
one has
\(
\Delta_{W_i}
<
(1-\chi_{W_i})
[
C_{W_i1}V_{W_i}^{\mu_i}
+
C_{W_i2}V_{W_i}^{\nu_i}
]
\),
which implies
\(
\dot V_{W_i}
\leq
-\chi_{W_i}C_{W_i1}V_{W_i}^{\mu_i}
-\chi_{W_i}C_{W_i2}V_{W_i}^{\nu_i}
\).
Since
\(
0<\mu_i<1<\nu_i
\),
Lemma~\ref{lem:fixed_time}
gives the uniform settling-time bound
\(
T_{W_i}
\leq
\frac{1}{\chi_{W_i}C_{W_i1}(1-\mu_i)}
+
\frac{1}{\chi_{W_i}C_{W_i2}(\nu_i-1)}
\),
independently of the initial critic-weight error.

At the boundary
\(
V_{W_i}=\bar V_{W_i}
\),
\eqref{eq:critic_fixed_time_residual_equation}
gives
\(
C_{W_i1}\bar V_{W_i}^{\mu_i}
+
C_{W_i2}\bar V_{W_i}^{\nu_i}
=
\Delta_{W_i}/(1-\chi_{W_i})
\).
Therefore,
\(
\dot V_{W_i}
\leq
-\frac{\Delta_{W_i}}{1-\chi_{W_i}}
+\Delta_{W_i}
=
-\frac{\chi_{W_i}}{1-\chi_{W_i}}
\Delta_{W_i}
\leq0
\),
establishing forward invariance of
\(
\mathcal B_{W_i}
\).

Finally,
\(
V_{W_i}
\geq
[2\lambda_{\max}(\bGamma_i)]^{-1}
\|\tilde{\bW}_i\|^2
\)
and
\(
V_{W_i}(t)\leq\bar V_{W_i}
\)
imply
\(
\|\tilde{\bW}_i(t)\|
\leq
\sqrt{
2\lambda_{\max}(\bGamma_i)\bar V_{W_i}
}
\)
for all
\(
t\geq T_{W_i}
\).
This completes the proof.
\end{proof}

\section{Practical Fixed-Time Leader--Follower Consensus}
\label{sec:network_stability}

For
\(
\vartheta_i>0
\),
\(
\lambda_i>0
\),
and
\(
\eta>0
\),
consider
\begin{equation}
\begin{aligned}
\mathcal J(t)
:={}&
\sum_{i=1}^{N}
\vartheta_iV_i^{*}(\be_i(t))
+
\sum_{i=1}^{N}
\lambda_iV_{W_i}(t)
\\
&+
\eta
\int_{t-T}^{t}
\left[
\sum_{i=1}^{N}
\vartheta_iV_i^{*}(\be_i(\tau))
+
\sum_{i=1}^{N}
\lambda_iV_{W_i}(\tau)
\right]d\tau ,
\end{aligned}
\label{eq:network_LKF}
\end{equation}
where
\(
T>0
\)
and
\(
V_{W_i}
=
\frac12
\tilde{\bW}_i^{T}
\bGamma_i^{-1}
\tilde{\bW}_i
\).
Since
\(
V_i^{*}(\be_i)>0
\)
for
\(
\be_i\neq\mathbf0
\)
and
\(
V_{W_i}>0
\)
for
\(
\tilde{\bW}_i\neq\mathbf0
\),
one has
\(
\mathcal J(t)>0
\)
for every nonzero synchronization or critic-weight error.


\begin{lemma}
\label{lem:derivative_limited_window_bound}

Let
\(
y:[t-T,t]\rightarrow\mathbb R_{\geq0}
\)
be absolutely continuous and satisfy
\(
|\dot y(\tau)|\leq L_y
\)
for almost every
\(
\tau\in[t-T,t]
\),
where
\(
L_y>0
\).
Assume that
\(
\int_{t-T}^{t}y(\tau)d\tau\leq\bar I
\)
for some
\(
\bar I>0
\).
For
\(
0<\gamma<1
\),
define
\(
\mu_\gamma:=(\gamma+1)/2
\).
Then
\begin{equation}
\int_{t-T}^{t}
y^\gamma(\tau)d\tau
\geq
\rho_{\gamma,T}
\left(
\int_{t-T}^{t}
y(\tau)d\tau
\right)^{\mu_\gamma},
\label{eq:window_low_power_lemma}
\end{equation}
where
\begin{equation}
\rho_{\gamma,T}
:=
\min
\left\{
2^{\gamma-1}
L_y^{\frac{\gamma-1}{2}},
\;
4^{\gamma-1}
T^{1-\gamma}
\bar I^{\frac{\gamma-1}{2}}
\right\}.
\label{eq:window_low_power_rho}
\end{equation}

\end{lemma}

\begin{proof}
    See Appendix~\ref{Movingwin}
\end{proof}

\begin{theorem}[Practical fixed-time leader--follower consensus]
\label{thm:network_fixed_time}

Suppose that
Assumptions~\ref{ass:graph},
\ref{ass:finite_excitation_i},
\ref{ass:local_inverse_optimal_feasibility},
and
\ref{ass:replay_ie}
hold, and let the critic weights be updated according to
\eqref{eq:critic_update}.
For each follower $i\in\mathcal V$, use the fixed-time exponents
\(
\mu_i=(\gamma_1+1)/2
\)
and
\(
\nu_i=(\gamma_2+1)/2
\),
with
\(1/2<\mu_i<1<\nu_i\),
as in Theorem~\ref{thm:critic_fixed_time}.
Since $\gamma_1$ and $\gamma_2$ are common to all followers,
the exponent values are identical throughout the network.
Assume further that
\[
\sum_{i=1}^{N}
\vartheta_iV_i^{*}(\be_i(t))
+
\sum_{i=1}^{N}
\lambda_iV_{W_i}(t)
\]
is absolutely continuous and satisfies
\[
\left|
\frac{d}{dt}
\left[
\sum_{i=1}^{N}
\vartheta_iV_i^{*}
+
\sum_{i=1}^{N}
\lambda_iV_{W_i}
\right]
\right|
\leq
L_{\Sigma}
\]
and
\[
\int_{t-T}^{t}
\left[
\sum_{i=1}^{N}
\vartheta_iV_i^{*}(\be_i(\tau))
+
\sum_{i=1}^{N}
\lambda_iV_{W_i}(\tau)
\right]d\tau
\leq
\bar I_{\Sigma}
\]
for finite
\(
L_{\Sigma}>0
\)
and
\(
\bar I_{\Sigma}>0
\).
Then there exist
\(
C_{\mathcal J1}>0
\),
\(
C_{\mathcal J2}>0
\),
and
\(
\Delta_{\mathcal J}\geq0
\)
such that

\begin{equation}
\begin{aligned}
\dot{\mathcal J}
\leq{}&
-C_{\mathcal J1}\mathcal J^{\bar\mu_i}
-C_{\mathcal J2}\mathcal J^{\nu_i}
+
\Delta_{\mathcal J},
\bar\mu_i
:=
\frac{\mu_i+1}{2},
\qquad
0<\bar\mu_i<1<\nu_i.
\end{aligned}
\label{eq:network_integral_FxT}
\end{equation}

Consequently, the synchronization and critic-weight errors converge in fixed time to a bounded residual set. If
\(
\Delta_{\mathcal J}=0
\),
exact fixed-time convergence is recovered.
\end{theorem}

\begin{proof} From
Theorem~\ref{thm:critic_fixed_time},
for every follower
\(
i
\),
\(
\dot V_{W_i}
\leq
-C_{W_i1}V_{W_i}^{\mu_i}
-C_{W_i2}V_{W_i}^{\nu_i}
+
\Delta_{W_i}
\).
Multiplying by
\(
\lambda_i>0
\)
and summing from
\(
i=1
\)
to
\(
N
\)
give
\(
\frac{d}{dt}
[
\sum_{i=1}^{N}
\lambda_iV_{W_i}
]
\leq
-
\sum_{i=1}^{N}
\lambda_iC_{W_i1}
V_{W_i}^{\mu_i}
-
\sum_{i=1}^{N}
\lambda_iC_{W_i2}
V_{W_i}^{\nu_i}
+
\sum_{i=1}^{N}
\lambda_i\Delta_{W_i}
\). Since
\(
\lambda_iV_{W_i}^{\mu_i}
=
\lambda_i^{1-\mu_i}
(\lambda_iV_{W_i})^{\mu_i}
\),
one obtains
\(
\sum_{i=1}^{N}
\lambda_iC_{W_i1}V_{W_i}^{\mu_i}
=
\sum_{i=1}^{N}
C_{W_i1}
\lambda_i^{1-\mu_i}
(\lambda_iV_{W_i})^{\mu_i}
\geq
\min_i
\{
C_{W_i1}
\lambda_i^{1-\mu_i}
\}
\sum_{i=1}^{N}
(\lambda_iV_{W_i})^{\mu_i}
\).
Because
\(
0<\mu_i<1
\),
the map
\(
s\mapsto s^{\mu_i}
\)
is concave and satisfies
\(
\sum_{i=1}^{N}
(\lambda_iV_{W_i})^{\mu_i}
\geq
[
\sum_{i=1}^{N}
\lambda_iV_{W_i}
]^{\mu_i}
\).
Therefore,
\(
\sum_{i=1}^{N}
\lambda_iC_{W_i1}V_{W_i}^{\mu_i}
\geq
C_{W1}
[
\sum_{i=1}^{N}
\lambda_iV_{W_i}
]^{\mu_i}
\),
where
\(
C_{W1}
:=
\min_i
\{
C_{W_i1}
\lambda_i^{1-\mu_i}
\}
\). For the high-order term,
\(
\lambda_iV_{W_i}^{\nu_i}
=
\lambda_i^{1-\nu_i}
(\lambda_iV_{W_i})^{\nu_i}
\),
hence
\(
\sum_{i=1}^{N}
\lambda_iC_{W_i2}V_{W_i}^{\nu_i}
=
\sum_{i=1}^{N}
C_{W_i2}
\lambda_i^{1-\nu_i}
(\lambda_iV_{W_i})^{\nu_i}
\geq
\min_i
\{
C_{W_i2}
\lambda_i^{1-\nu_i}
\}
\sum_{i=1}^{N}
(\lambda_iV_{W_i})^{\nu_i}
\).
Since
\(
\nu_i>1
\),
convexity gives
\(
[
\frac1N
\sum_{i=1}^{N}
\lambda_iV_{W_i}
]^{\nu_i}
\leq
\frac1N
\sum_{i=1}^{N}
(\lambda_iV_{W_i})^{\nu_i}
\),
or equivalently,
\(
\sum_{i=1}^{N}
(\lambda_iV_{W_i})^{\nu_i}
\geq
N^{1-\nu_i}
[
\sum_{i=1}^{N}
\lambda_iV_{W_i}
]^{\nu_i}
\).
Thus,
\(
\sum_{i=1}^{N}
\lambda_iC_{W_i2}V_{W_i}^{\nu_i}
\geq
C_{W2}
[
\sum_{i=1}^{N}
\lambda_iV_{W_i}
]^{\nu_i}
\),
where
\(
C_{W2}
:=
N^{1-\nu_i}
\min_i
\{
C_{W_i2}
\lambda_i^{1-\nu_i}
\}
\).
Defining
\(
\Delta_W
:=
\sum_{i=1}^{N}
\lambda_i\Delta_{W_i}
\),
one obtains

\begin{equation}
\begin{aligned}
\frac{d}{dt}
\left[
\sum_{i=1}^{N}\lambda_i V_{W_i}
\right]
\leq{}&
-C_{W1}
\left[
\sum_{i=1}^{N}\lambda_i V_{W_i}
\right]^{\mu_i}
\\
&-C_{W2}
\left[
\sum_{i=1}^{N}\lambda_i V_{W_i}
\right]^{\nu_i}
+\Delta_W.
\end{aligned}
\label{eq:aggregate_critic_direct}
\end{equation}


For the synchronization dynamics, the critic-generated and ideal local policies satisfy
\(
\hat{\bar{\bu}}_i
-
\bar{\bu}_i^{*}
=
-\frac12
\bar{\bR}_i^{-1}
\bar{\mathbf G}_i^T
(
\nabla\hat V_i
-
\nabla V_i^{*}
)
\).
Using
\(
V_i^{*}
=
\bW_i^{*T}\bphi_i
+
\varepsilon_i
\),
\(
\hat V_i
=
\hat{\bW}_i^T\bphi_i
\),
and
\(
\tilde{\bW}_i
=
\bW_i^{*}
-
\hat{\bW}_i
\),
one has
\(
\nabla\hat V_i
-
\nabla V_i^{*}
=
-
(\nabla\bphi_i)^T
\tilde{\bW}_i
-
\nabla\varepsilon_i
\).
Hence,
\(
\hat{\bar{\bu}}_i
-
\bar{\bu}_i^{*}
=
\frac12
\bar{\bR}_i^{-1}
\bar{\mathbf G}_i^T
[
(\nabla\bphi_i)^T
\tilde{\bW}_i
+
\nabla\varepsilon_i
]
\).

On the compact operating set, let
\(
\|\bar{\mathbf G}_i\|
\leq
\bar G_i
\),
\(
\|\nabla\bphi_i\|
\leq
\bar\phi_{g_i}
\),
and
\(
\|\nabla\varepsilon_i\|
\leq
\bar\varepsilon_{g_i}
\).
Then
\(
\|
\hat{\bar{\bu}}_i
-
\bar{\bu}_i^{*}
\|
\leq
\frac12
\|\bar{\bR}_i^{-1}\|
\bar G_i
\bar\phi_{g_i}
\|\tilde{\bW}_i\|
+
\frac12
\|\bar{\bR}_i^{-1}\|
\bar G_i
\bar\varepsilon_{g_i}
\).

Since
\(
V_{W_i}
=
\frac12
\tilde{\bW}_i^T
\bGamma_i^{-1}
\tilde{\bW}_i
\geq
\frac{1}
{2\lambda_{\max}(\bGamma_i)}
\|\tilde{\bW}_i\|^2
\),
one has
\(
\|\tilde{\bW}_i\|
\leq
[
2\lambda_{\max}(\bGamma_i)
]^{1/2}
V_{W_i}^{1/2}
\).
Therefore,
\begin{equation}
\|
\hat{\bar{\bu}}_i
-
\bar{\bu}_i^{*}
\|
\leq
c_{\pi_i}V_{W_i}^{1/2}
+
\Delta_{u_i},
\label{eq:policy_error_local}
\end{equation}
where
\(
c_{\pi_i}
:=
\frac12
\|\bar{\bR}_i^{-1}\|
\bar G_i
\bar\phi_{g_i}
[
2\lambda_{\max}(\bGamma_i)
]^{1/2}
\)
and
\(
\Delta_{u_i}
:=
\frac12
\|\bar{\bR}_i^{-1}\|
\bar G_i
\bar\varepsilon_{g_i}
\).


By Lemma~\ref{lem:bounded_identification_error_i},
for each follower
\(
\ell\in\mathcal V
\),
\(
\|
\mathbf f_\ell(\bx_\ell)
-
\hat{\mathbf f}_\ell(\bx_\ell)
\|
\leq
\bar\Delta_{f_\ell}
\).
From
\(
\mathbf F_i^e
-
\hat{\mathbf F}_i^e
=
\kappa_i
[
\mathbf f_i(\bx_i)
-
\hat{\mathbf f}_i(\bx_i)
]
-
\sum_{j\in\mathcal N_i}
a_{ij}
[
\mathbf f_j(\bx_j)
-
\hat{\mathbf f}_j(\bx_j)
]
\),
the triangle inequality gives
\(
\|
\mathbf F_i^e
-
\hat{\mathbf F}_i^e
\|
\leq
\kappa_i
\|
\mathbf f_i-\hat{\mathbf f}_i
\|
+
\sum_{j\in\mathcal N_i}
a_{ij}
\|
\mathbf f_j-\hat{\mathbf f}_j
\|
\leq
\kappa_i\bar\Delta_{f_i}
+
\sum_{j\in\mathcal N_i}
a_{ij}\bar\Delta_{f_j}
\).
Hence,
\begin{equation}
\|
\mathbf F_i^e
-
\hat{\mathbf F}_i^e
\|
\leq
\Delta_{e_i},
\qquad
\Delta_{e_i}
:=
\kappa_i\bar\Delta_{f_i}
+
\sum_{j\in\mathcal N_i}
a_{ij}\bar\Delta_{f_j}.
\label{eq:network_identification_error_i}
\end{equation}


Under the learned distributed policy,
\(
\dot{\be}_i
=
\hat{\mathbf F}_i^e
+
\bar{\mathbf G}_i
\bar{\bu}_i^{*}
+
\bar{\mathbf G}_i
(
\hat{\bar{\bu}}_i
-
\bar{\bu}_i^{*}
)
+
(
\mathbf F_i^e
-
\hat{\mathbf F}_i^e
)
\).
Hence,
\(
\dot V_i^{*}
=
\nabla V_i^{*T}
[
\hat{\mathbf F}_i^e
+
\bar{\mathbf G}_i
\bar{\bu}_i^{*}
]
+
\nabla V_i^{*T}
\bar{\mathbf G}_i
(
\hat{\bar{\bu}}_i
-
\bar{\bu}_i^{*}
)
+
\nabla V_i^{*T}
(
\mathbf F_i^e
-
\hat{\mathbf F}_i^e
)
\).

By
Theorem~\ref{eq:identified_local_error_dynamics_compact},
\[
\nabla V_i^{*T}
\left[
\hat{\mathbf F}_i^e
+
\bar{\mathbf G}_i
\bar{\bu}_i^{*}
\right]
=
-a_i(V_i^{*})^{\mu_i}
-b_i(V_i^{*})^{\nu_i}.
\]
Let
\(
\|\nabla V_i^{*}\|
\leq
\bar V_{g_i}
\).
Then
\[
\begin{aligned}
\nabla V_i^{*T}
\bar{\mathbf G}_i
\left(
\hat{\bar{\bu}}_i
-
\bar{\bu}_i^{*}
\right)
&\leq
\bar V_{g_i}
\bar G_i
\|
\hat{\bar{\bu}}_i
-
\bar{\bu}_i^{*}
\|
\leq
\bar V_{g_i}
\bar G_i
c_{\pi_i}
V_{W_i}^{1/2}
+
\bar V_{g_i}
\bar G_i
\Delta_{u_i},
\end{aligned}
\]
while
\[
\nabla V_i^{*T}
\left(
\mathbf F_i^e
-
\hat{\mathbf F}_i^e
\right)
\leq
\bar V_{g_i}\Delta_{e_i}.
\]
Therefore,
\[
\dot V_i^{*}
\leq
-a_i(V_i^{*})^{\mu_i}
-b_i(V_i^{*})^{\nu_i}
+
c_{c_i}V_{W_i}^{1/2}
+
\Delta_{V_i},
\]
where
\[
c_{c_i}
:=
\bar V_{g_i}\bar G_i c_{\pi_i},
\qquad
\Delta_{V_i}
:=
\bar V_{g_i}
\left(
\bar G_i\Delta_{u_i}
+
\Delta_{e_i}
\right).
\]


\begin{lemma}[Absorption of the critic-induced coupling term]
\label{lem:critic_coupling_absorption}
For each follower $i\in\mathcal V$, suppose that
$1/2<\mu_i<1$ and $c_{c_i}\geq0$.
Then, for any $\varepsilon_i>0$ and $s\geq0$,
\(
c_{c_i}s
\leq
\varepsilon_i s^{2\mu_i}
+
\Delta_{c_i},
\)
where
\(
\Delta_{c_i}
:=
\frac{2\mu_i-1}{2\mu_i}
c_{c_i}
\left(
\frac{c_{c_i}}
{2\mu_i\varepsilon_i}
\right)^{\frac{1}{2\mu_i-1}}
\).
\end{lemma}

\begin{proof}
    See Appendix~\ref{Appen1}
\end{proof}

Applying
Lemma~\ref{lem:critic_coupling_absorption}
with
\(
s=V_{W_i}^{1/2}
\)
is admissible since
\(
V_{W_i}\geq0
\).
Noting that
\(
s^{2\mu_i}
=
\left(V_{W_i}^{1/2}\right)^{2\mu_i}
=
V_{W_i}^{\mu_i}
\),
one obtains
\(
c_{c_i}V_{W_i}^{1/2}
\leq
\varepsilon_iV_{W_i}^{\mu_i}
+
\Delta_{c_i}
\).
Substituting this bound into the preceding
Lyapunov derivative inequality yields
\begin{equation}
\dot V_i^{*}
\leq
-a_i(V_i^{*})^{\mu_i}
-b_i(V_i^{*})^{\nu_i}
+
\varepsilon_iV_{W_i}^{\mu_i}
+
\Delta_{V_i}
+
\Delta_{c_i}.
\label{eq:local_state_absorbed}
\end{equation}

Multiplying
\eqref{eq:local_state_absorbed}
by
\(
\vartheta_i
\)
and summing over
\(
i=1,\ldots,N
\)
give
\(
\frac{d}{dt}
\left[
\sum_{i=1}^{N}
\vartheta_iV_i^{*}
\right]
\leq
-
\sum_{i=1}^{N}
\vartheta_ia_i
(V_i^{*})^{\mu_i}
-
\sum_{i=1}^{N}
\vartheta_ib_i
(V_i^{*})^{\nu_i}
+
\sum_{i=1}^{N}
\vartheta_i\varepsilon_i
V_{W_i}^{\mu_i}
+
\Delta_d
\),
where
\(
\Delta_d
:=
\sum_{i=1}^{N}
\vartheta_i
\left(
\Delta_{V_i}
+
\Delta_{c_i}
\right)
\).



Now, for nonnegative scalars $x_i$, the power-sum
inequalities give
\(
\sum_{i=1}^{N}x_i^{p}
\geq
(\sum_{i=1}^{N}x_i)^{p}
\)
for
\(
0<p<1
\),
and
\(
\sum_{i=1}^{N}x_i^{p}
\geq
N^{1-p}(\sum_{i=1}^{N}x_i)^{p}
\)
for
\(
p>1
\).
Applying the first inequality with
\(
x_i=\vartheta_iV_i^{*}
\)
and the common exponent
\(
p=\mu_i
\)
yields
\[
\begin{aligned}
\sum_{i=1}^{N}
\vartheta_ia_i
(V_i^{*})^{\mu_i}
&=
\sum_{i=1}^{N}
a_i
\vartheta_i^{1-\mu_i}
(\vartheta_iV_i^{*})^{\mu_i}\\
&\geq
\min_i
\left\{
a_i\vartheta_i^{1-\mu_i}
\right\}
\sum_{i=1}^{N}
(\vartheta_iV_i^{*})^{\mu_i}
\geq
A_1
\left[
\sum_{i=1}^{N}
\vartheta_iV_i^{*}
\right]^{\mu_i},
\end{aligned}
\]
where
\(A_1 := \min_i \left\{ a_i\vartheta_i^{1-\mu_i} \right\}.\)

Similarly, applying the second power-sum inequality
with the common exponent
\(
p=\nu_i>1
\)
gives
\[
\begin{aligned}
\sum_{i=1}^{N}
\vartheta_ib_i
(V_i^{*})^{\nu_i}
&=
\sum_{i=1}^{N}
b_i
\vartheta_i^{1-\nu_i}
(\vartheta_iV_i^{*})^{\nu_i}\\
&\geq
\min_i
\left\{
b_i\vartheta_i^{1-\nu_i}
\right\}
\sum_{i=1}^{N}
(\vartheta_iV_i^{*})^{\nu_i}
\geq
A_2
\left[
\sum_{i=1}^{N}
\vartheta_iV_i^{*}
\right]^{\nu_i},
\end{aligned}
\]
where
\(A_2 := N^{1-\nu_i} \min_i \left\{ b_i\vartheta_i^{1-\nu_i} \right\}.\)

Furthermore,
\[
\begin{aligned}
\sum_{i=1}^{N}
\vartheta_i\varepsilon_i
V_{W_i}^{\mu_i}
&=
\sum_{i=1}^{N}
\vartheta_i\varepsilon_i
\lambda_i^{-\mu_i}
(\lambda_iV_{W_i})^{\mu_i}
\leq
\max_i
\left\{
\vartheta_i\varepsilon_i
\lambda_i^{-\mu_i}
\right\}
\sum_{i=1}^{N}
(\lambda_iV_{W_i})^{\mu_i}.
\end{aligned}
\]
For
\(
0<\mu_i<1
\),
concavity gives
\(\sum_{i=1}^{N} (\lambda_iV_{W_i})^{\mu_i} \leq N^{1-\mu_i} \left[ \sum_{i=1}^{N} \lambda_iV_{W_i} \right]^{\mu_i}.\)

Therefore,
\(\sum_{i=1}^{N} \vartheta_i\varepsilon_i V_{W_i}^{\mu_i} \leq E_W \left[ \sum_{i=1}^{N} \lambda_iV_{W_i} \right]^{\mu_i},\)
where
\(E_W := N^{1-\mu_i} \max_i \left\{ \vartheta_i\varepsilon_i \lambda_i^{-\mu_i} \right\}.\)

Hence,
\begin{equation}
\begin{aligned}
\frac{d}{dt}
\left[
\sum_{i=1}^{N}
\vartheta_iV_i^{*}
\right]
\leq{}&
-A_1
\left[
\sum_{i=1}^{N}
\vartheta_iV_i^{*}
\right]^{\mu_i}
-
A_2
\left[
\sum_{i=1}^{N}
\vartheta_iV_i^{*}
\right]^{\nu_i}
\\
&+
E_W
\left[
\sum_{i=1}^{N}
\lambda_iV_{W_i}
\right]^{\mu_i}
+
\Delta_d.
\end{aligned}
\label{eq:aggregate_state_direct}
\end{equation}

Choose
\(
\varepsilon_i
\)
such that
\(
E_W<C_{W1}
\),
and define
\(\bar C_{W1} := C_{W1}-E_W >0.\)
Adding
\eqref{eq:aggregate_critic_direct}
and
\eqref{eq:aggregate_state_direct}
gives
\[
\begin{aligned}
\frac{d}{dt}
\Bigg[
&
\sum_{i=1}^{N}
\vartheta_iV_i^{*}
+
\sum_{i=1}^{N}
\lambda_iV_{W_i}
\Bigg]
\leq{}
-A_1
\left[
\sum_{i=1}^{N}
\vartheta_iV_i^{*}
\right]^{\mu_i}\\
&-
\bar C_{W1}
\left[
\sum_{i=1}^{N}
\lambda_iV_{W_i}
\right]^{\mu_i}
-
A_2
\left[
\sum_{i=1}^{N}
\vartheta_iV_i^{*}
\right]^{\nu_i}\\
&-
C_{W2}
\left[
\sum_{i=1}^{N}
\lambda_iV_{W_i}
\right]^{\nu_i}
+
\Delta_{\Sigma},
\end{aligned}
\]
where
\(\Delta_{\Sigma} := \Delta_d+\Delta_W.\)

Since
\[
\left[
\sum_{i=1}^{N}
\vartheta_iV_i^{*}
\right]^{\mu_i}
+
\left[
\sum_{i=1}^{N}
\lambda_iV_{W_i}
\right]^{\mu_i}
\geq
\left[
\sum_{i=1}^{N}
\vartheta_iV_i^{*}
+
\sum_{i=1}^{N}
\lambda_iV_{W_i}
\right]^{\mu_i},
\]
one has
\[
\begin{aligned}
&A_1
\left[
\sum_{i=1}^{N}
\vartheta_iV_i^{*}
\right]^{\mu_i}
+
\bar C_{W1}
\left[
\sum_{i=1}^{N}
\lambda_iV_{W_i}
\right]^{\mu_i}
\\
&\geq
C_{\Sigma1}
\left[
\sum_{i=1}^{N}
\vartheta_iV_i^{*}
+
\sum_{i=1}^{N}
\lambda_iV_{W_i}
\right]^{\mu_i},
\end{aligned}
\]
where
\(C_{\Sigma1} := \min \left\{ A_1,\bar C_{W1} \right\}.\)

Likewise,
\[
\begin{aligned}
&A_2
\left[
\sum_{i=1}^{N}
\vartheta_iV_i^{*}
\right]^{\nu_i}
+
C_{W2}
\left[
\sum_{i=1}^{N}
\lambda_iV_{W_i}
\right]^{\nu_i}\geq\\
&C_{\Sigma2}
\left[
\sum_{i=1}^{N}
\vartheta_iV_i^{*}
+
\sum_{i=1}^{N}
\lambda_iV_{W_i}
\right]^{\nu_i},
\end{aligned}
\]
where
\(C_{\Sigma2} := 2^{1-\nu_i} \min \left\{ A_2,C_{W2} \right\}.\)

Therefore,
\begin{equation}
\begin{aligned}
&\frac{d}{dt}
\left(
\sum_{i=1}^{N}\vartheta_iV_i^{*}
+
\sum_{i=1}^{N}\lambda_iV_{W_i}
\right)
\leq
-C_{\Sigma1}
\left(
\sum_{i=1}^{N}\vartheta_iV_i^{*}
+
\sum_{i=1}^{N}\lambda_iV_{W_i}
\right)^{\mu_i}
\\
&\quad
-C_{\Sigma2}
\left(
\sum_{i=1}^{N}\vartheta_iV_i^{*}
+
\sum_{i=1}^{N}\lambda_iV_{W_i}
\right)^{\nu_i}
+
\Delta_{\Sigma}.
\end{aligned}
\label{eq:pointwise_full_network}
\end{equation}

Differentiating the finite-window term in
\eqref{eq:network_LKF}
gives
\[
\begin{aligned}
&\frac{d}{dt}
\int_{t-T}^{t}
\left[
\sum_{i=1}^{N}
\vartheta_iV_i^{*}(\be_i(\tau))
+
\sum_{i=1}^{N}
\lambda_iV_{W_i}(\tau)
\right]d\tau
\\
&=
\left[
\sum_{i=1}^{N}
\vartheta_iV_i^{*}(\be_i(t))
+
\sum_{i=1}^{N}
\lambda_iV_{W_i}(t)
\right]
\\
&\quad-
\left[
\sum_{i=1}^{N}
\vartheta_iV_i^{*}(\be_i(t-T))
+
\sum_{i=1}^{N}
\lambda_iV_{W_i}(t-T)
\right].
\end{aligned}
\]
By absolute continuity,
\[
\begin{aligned}
&\left[
\sum_{i=1}^{N}
\vartheta_iV_i^{*}(\be_i(t))
+
\sum_{i=1}^{N}
\lambda_iV_{W_i}(t)
\right]
\\
&\quad-
\left[
\sum_{i=1}^{N}
\vartheta_iV_i^{*}(\be_i(t-T))
+
\sum_{i=1}^{N}
\lambda_iV_{W_i}(t-T)
\right]
\\
&=
\int_{t-T}^{t}
\frac{d}{d\tau}
\left[
\sum_{i=1}^{N}
\vartheta_iV_i^{*}(\be_i(\tau))
+
\sum_{i=1}^{N}
\lambda_iV_{W_i}(\tau)
\right]d\tau.
\end{aligned}
\]

Using
\eqref{eq:pointwise_full_network},
\[
\begin{aligned}
&\frac{d}{dt}
\int_{t-T}^{t}
\left[
\sum_{i=1}^{N}
\vartheta_iV_i^{*}
+
\sum_{i=1}^{N}
\lambda_iV_{W_i}
\right]d\tau
\\
&\leq
-C_{\Sigma1}
\int_{t-T}^{t}
\left[
\sum_{i=1}^{N}
\vartheta_iV_i^{*}(\tau)
+
\sum_{i=1}^{N}
\lambda_iV_{W_i}(\tau)
\right]^{\mu_i}
d\tau
\\
&\quad-
C_{\Sigma2}
\int_{t-T}^{t}
\left[
\sum_{i=1}^{N}
\vartheta_iV_i^{*}(\tau)
+
\sum_{i=1}^{N}
\lambda_iV_{W_i}(\tau)
\right]^{\nu_i}
d\tau
+
T\Delta_{\Sigma}.
\end{aligned}
\]

Now apply
Lemma~\ref{lem:derivative_limited_window_bound}
with
\(y(\tau) = \sum_{i=1}^{N} \vartheta_iV_i^{*}(\be_i(\tau)) + \sum_{i=1}^{N} \lambda_iV_{W_i}(\tau), \qquad \gamma=\mu_i.\)
Since
\(
0<\mu_i<1
\),
Lemma~\ref{lem:derivative_limited_window_bound}
gives
\[
\begin{aligned}
&\int_{t-T}^{t}
\left[
\sum_{i=1}^{N}
\vartheta_iV_i^{*}(\tau)
+
\sum_{i=1}^{N}
\lambda_iV_{W_i}(\tau)
\right]^{\mu_i}
d\tau
\\
&\geq
\rho_{\mu_i,T}
\left\{
\int_{t-T}^{t}
\left[
\sum_{i=1}^{N}
\vartheta_iV_i^{*}(\tau)
+
\sum_{i=1}^{N}
\lambda_iV_{W_i}(\tau)
\right]d\tau
\right\}^{\bar\mu_i},
\end{aligned}
\]
where
\(\bar\mu_i = \frac{\mu_i+1}{2}\)
and
\[
\rho_{\mu_i,T}
=
\min
\left\{
2^{\mu_i-1}
L_{\Sigma}^{\frac{\mu_i-1}{2}},
\;
4^{\mu_i-1}
T^{1-\mu_i}
\bar I_{\Sigma}^{\frac{\mu_i-1}{2}}
\right\}.
\]

For
\(
\nu_i>1
\),
the function
\(
s\mapsto s^{\nu_i}
\)
is convex.
Therefore, Jensen's inequality yields
\(\left[ \frac1T \int_{t-T}^{t} \left( \sum_{i=1}^{N} \vartheta_iV_i^{*}(\tau) + \sum_{i=1}^{N} \lambda_iV_{W_i}(\tau) \right)d\tau \right]^{\nu_i}\)
\(\leq \frac1T \int_{t-T}^{t} \left[ \sum_{i=1}^{N} \vartheta_iV_i^{*}(\tau) + \sum_{i=1}^{N} \lambda_iV_{W_i}(\tau) \right]^{\nu_i} d\tau.\)
Hence,
\[
\begin{aligned}
&\int_{t-T}^{t}
\left[
\sum_{i=1}^{N}
\vartheta_iV_i^{*}(\tau)
+
\sum_{i=1}^{N}
\lambda_iV_{W_i}(\tau)
\right]^{\nu_i}
d\tau
\\
&\geq
T^{1-\nu_i}
\left\{
\int_{t-T}^{t}
\left[
\sum_{i=1}^{N}
\vartheta_iV_i^{*}(\tau)
+
\sum_{i=1}^{N}
\lambda_iV_{W_i}(\tau)
\right]d\tau
\right\}^{\nu_i}.
\end{aligned}
\]

Therefore,


\[
\begin{aligned}
\dot{\mathcal J}
\leq{}&
-C_{\Sigma1}
\Bigl[
\sum_{i=1}^{N}\vartheta_iV_i^{*}
+
\sum_{i=1}^{N}\lambda_iV_{W_i}
\Bigr]^{\mu_i}
-
C_{\Sigma2}
\Bigl[
\sum_{i=1}^{N}\vartheta_iV_i^{*}
+
\sum_{i=1}^{N}\lambda_iV_{W_i}
\Bigr]^{\nu_i}
\\
&-
\eta C_{\Sigma1}\rho_{\mu_i,T}
\Biggl\{
\int_{t-T}^{t}
\Bigl[
\sum_{i=1}^{N}\vartheta_iV_i^{*}(\tau)
+
\sum_{i=1}^{N}\lambda_iV_{W_i}(\tau)
\Bigr]d\tau
\Biggr\}^{\bar\mu_i}
\\
&-
\eta C_{\Sigma2}T^{1-\nu_i}
\Biggl\{
\int_{t-T}^{t}
\Bigl[
\sum_{i=1}^{N}\vartheta_iV_i^{*}(\tau)
+
\sum_{i=1}^{N}\lambda_iV_{W_i}(\tau)
\Bigr]d\tau
\Biggr\}^{\nu_i}
\\
&+
(1+\eta T)\Delta_{\Sigma}.
\end{aligned}
\]

On the compact operating domain, let
\[
\begin{aligned}
&\sum_{i=1}^{N}\vartheta_iV_i^{*}
+
\sum_{i=1}^{N}\lambda_iV_{W_i}
\leq \bar V_{\Sigma}.
\end{aligned}
\]
Since
\(
\bar\mu_i=(\mu_i+1)/2>\mu_i
\),
one has
\[
\begin{aligned}
&
\Bigl[
\sum_{i=1}^{N}\vartheta_iV_i^{*}
+
\sum_{i=1}^{N}\lambda_iV_{W_i}
\Bigr]^{\mu_i}
=
\Bigl[
\sum_{i=1}^{N}\vartheta_iV_i^{*}
+
\sum_{i=1}^{N}\lambda_iV_{W_i}
\Bigr]^{\bar\mu_i}
\\
&\quad\times
\Bigl[
\sum_{i=1}^{N}\vartheta_iV_i^{*}
+
\sum_{i=1}^{N}\lambda_iV_{W_i}
\Bigr]^{\mu_i-\bar\mu_i}
\\
&\geq
\bar V_{\Sigma}^{\mu_i-\bar\mu_i}
\Bigl[
\sum_{i=1}^{N}\vartheta_iV_i^{*}
+
\sum_{i=1}^{N}\lambda_iV_{W_i}
\Bigr]^{\bar\mu_i}.
\end{aligned}
\]

Moreover,
\[
\begin{aligned}
&\eta
\left\{
\int_{t-T}^{t}
\left[
\sum_{i=1}^{N}
\vartheta_iV_i^{*}(\tau)
+
\sum_{i=1}^{N}
\lambda_iV_{W_i}(\tau)
\right]d\tau
\right\}^{\bar\mu_i}
\\
&=
\eta^{1-\bar\mu_i}
\left\{
\eta
\int_{t-T}^{t}
\left[
\sum_{i=1}^{N}
\vartheta_iV_i^{*}(\tau)
+
\sum_{i=1}^{N}
\lambda_iV_{W_i}(\tau)
\right]d\tau
\right\}^{\bar\mu_i}.
\end{aligned}
\]

Define
\(\widehat C_1 := \min \left\{ C_{\Sigma1} \bar V_{\Sigma}^{\mu_i-\bar\mu_i}, \; C_{\Sigma1} \rho_{\mu_i,T} \eta^{1-\bar\mu_i} \right\}.\)
Since
\(
0<\bar\mu_i<1
\),
for any
\(
a,b\geq0
\),
\(
a^{\bar\mu_i}+b^{\bar\mu_i}\geq(a+b)^{\bar\mu_i}
\).
Hence,
\[
\begin{aligned}
&
C_{\Sigma1}
\left[
\sum_{i=1}^{N}
\vartheta_iV_i^{*}
+
\sum_{i=1}^{N}
\lambda_iV_{W_i}
\right]^{\mu_i}
\\
&+
\eta C_{\Sigma1}\rho_{\mu_i,T}
\left\{
\int_{t-T}^{t}
\left[
\sum_{i=1}^{N}
\vartheta_iV_i^{*}(\tau)
+
\sum_{i=1}^{N}
\lambda_iV_{W_i}(\tau)
\right]d\tau
\right\}^{\bar\mu_i}
\\
&\geq
\widehat C_1
\mathcal J^{\bar\mu_i}.
\end{aligned}
\]

For the high-order part,

\[
\begin{aligned}
&\eta C_{\Sigma2}T^{1-\nu_i}
\Biggl\{
\int_{t-T}^{t}
\Bigl[
\sum_{i=1}^{N}\vartheta_iV_i^{*}(\tau)
+\sum_{i=1}^{N}\lambda_iV_{W_i}(\tau)
\Bigr]d\tau
\Biggr\}^{\nu_i}
\\
&=
C_{\Sigma2}T^{1-\nu_i}\eta^{1-\nu_i}
\Biggl\{
\eta\int_{t-T}^{t}
\Bigl[
\sum_{i=1}^{N}\vartheta_iV_i^{*}(\tau)
+\sum_{i=1}^{N}\lambda_iV_{W_i}(\tau)
\Bigr]d\tau
\Biggr\}^{\nu_i}.
\end{aligned}
\]

Define
\(\widehat C_2 := \min \left\{ C_{\Sigma2}, \; C_{\Sigma2} T^{1-\nu_i} \eta^{1-\nu_i} \right\}.\)
Since
\(
\nu_i>1
\),
for any
\(
a,b\geq0
\),
\(a^{\nu_i}+b^{\nu_i} \geq 2^{1-\nu_i}(a+b)^{\nu_i}.\)
Therefore,
\[
\begin{aligned}
&
C_{\Sigma2}
\left[
\sum_{i=1}^{N}
\vartheta_iV_i^{*}
+
\sum_{i=1}^{N}
\lambda_iV_{W_i}
\right]^{\nu_i}
\\
&+
\eta C_{\Sigma2}T^{1-\nu_i}
\left\{
\int_{t-T}^{t}
\left[
\sum_{i=1}^{N}
\vartheta_iV_i^{*}(\tau)
+
\sum_{i=1}^{N}
\lambda_iV_{W_i}(\tau)
\right]d\tau
\right\}^{\nu_i}
\\
&\geq
2^{1-\nu_i}
\widehat C_2
\mathcal J^{\nu_i}.
\end{aligned}
\]

Define
\(
C_{\mathcal J1}
:=
\widehat C_1
\),
\(
C_{\mathcal J2}
:=
2^{1-\nu_i}
\widehat C_2
\),
and
\(
\Delta_{\mathcal J}
:=
(1+\eta T)\Delta_{\Sigma}
\).
It follows that
\(
\dot{\mathcal J}
\leq
-C_{\mathcal J1}
\mathcal J^{\bar\mu_i}
-C_{\mathcal J2}
\mathcal J^{\nu_i}
+
\Delta_{\mathcal J}
\).

For any
\(
0<\chi<1
\),
Lemma~\ref{lem:fixed_time}
applied with
\(
a=C_{\mathcal J1}
\),
\(
b=C_{\mathcal J2}
\),
\(
\mu_i=\bar\mu_i
\),
\(
\nu_i=\nu_i
\),
and
\(
\Delta=\Delta_{\mathcal J}
\)
implies that
\(
\Omega_{\mathcal J,\chi}
:=
\{
\mathcal J\geq0:
C_{\mathcal J1}\mathcal J^{\bar\mu_i}
+
C_{\mathcal J2}\mathcal J^{\nu_i}
\leq
\Delta_{\mathcal J}/(1-\chi)
\}
\)
is reached within
\begin{equation}
T_{\mathcal J}
\leq
\frac{1}
{\chi C_{\mathcal J1}(1-\bar\mu_i)}
+
\frac{1}
{\chi C_{\mathcal J2}(\nu_i-1)}.
\label{eq:network_fixed_time}
\end{equation}

Let
\(
\mathcal J_{\Delta}
\)
be
\(
C_{\mathcal J1}
\mathcal J_{\Delta}^{\bar\mu_i}
+
C_{\mathcal J2}
\mathcal J_{\Delta}^{\nu_i}
=
\Delta_{\mathcal J}/(1-\chi)
\).
Since

\(
\sum_{i=1}^{N}
\vartheta_iV_i^{*}(\be_i)
\leq
\mathcal J
\)
and, on the compact synchronization domain,
\(
\sum_{i=1}^{N}
\vartheta_iV_i^{*}(\be_i)
\geq
\underline\alpha_e
\|\be\|^2
\)
for some
\(
\underline\alpha_e>0
\),
one obtains, for
\(
t\geq T_{\mathcal J}
\),
\(
\underline\alpha_e
\|\be(t)\|^2
\leq
\mathcal J_\Delta
\),
and hence
\(
\|\be(t)\|
\leq
\sqrt{
\mathcal J_\Delta/
\underline\alpha_e
}
\).

Moreover,
\(
\lambda_iV_{W_i}
=
\frac{\lambda_i}{2}
\tilde{\bW}_i^T
\bGamma_i^{-1}
\tilde{\bW}_i
\leq
\mathcal J_\Delta
\).
Since
\(
V_{W_i}
\geq
\frac{1}
{2\lambda_{\max}(\bGamma_i)}
\|\tilde{\bW}_i\|^2
\),
one obtains
\(
\|\tilde{\bW}_i(t)\|
\leq
\sqrt{
\frac{
2\lambda_{\max}(\bGamma_i)
\mathcal J_\Delta
}{
\lambda_i
}
}
\),
\(
i=1,\ldots,N
\).

Finally, define \(\be:=\operatorname{col}(\be_1,\ldots,\be_N)\), one has
\(
\be
=
(\mathcal H\otimes I_n)
\tilde{\bx}
\),
and, by
Assumption~\ref{ass:graph},
\(
\mathcal H
\)
is nonsingular.
Thus,
\(
\tilde{\bx}
=
(\mathcal H^{-1}\otimes I_n)\be
\)
and
\(
\|\tilde{\bx}(t)\|
\leq
\|\mathcal H^{-1}\|
\sqrt{
\mathcal J_\Delta/
\underline\alpha_e
}
\).
Therefore, practical fixed-time leader--follower consensus is achieved.

If
\(
\Delta_{f_i}=0
\),
\(
\Delta_{u_i}=0
\),
and
\(
\Delta_{W_i}=0
\)
for every
\(
i
\),
then
\(
\Delta_{\Sigma}=0
\)
and
\(
\Delta_{\mathcal J}=0
\).
Hence,
\(
\dot{\mathcal J}
\leq
-C_{\mathcal J1}
\mathcal J^{\bar\mu_i}
-C_{\mathcal J2}
\mathcal J^{\nu_i}
\).
By
Lemma~\ref{lem:fixed_time},
\(
T_{\mathcal J}^{0}
\leq
1/
[
C_{\mathcal J1}(1-\bar\mu_i)
]
+
1/
[
C_{\mathcal J2}(\nu_i-1)
]
\),
which recovers exact fixed-time convergence.

\end{proof}

\subsection{Simulation Setup}

Consider a nonlinear leader--follower multi-agent
system consisting of one leader and four followers.
Each follower evolves in three-dimensional space
according to
\(
\dot{\bx}_i
=
\mathbf f_i(\bx_i)
+
\bg_i(\bx_i)\bu_i
\),
where
\(
\bx_i,\bu_i\in\mathbb R^3
\),
\(
\bg_i(\bx_i)=\bm I_3
\),
and \(i=1,\ldots,4\).
The unknown follower drifts are selected identically
as
\[
\mathbf f_i(\bx_i)=
\begin{bmatrix}
-\tanh(x_{i1}+x_{i2})-0.4x_{i1}^3\\
-\sin(x_{i2}+x_{i3})-0.3x_{i1}x_{i2}\\
-\dfrac{x_{i3}}{1+x_{i3}^2}
-0.2\tanh(x_{i1}x_{i3})
\end{bmatrix}.
\]


The leader follows a prescribed bounded
three-dimensional helical trajectory
\(
\bx_0(t)=
[
1.65(1-e^{-0.006t}),
\;
0.17\cos(0.1t),
\;
0.198(1-e^{-0.006t})
+0.17\sin(0.1t)
]^\top
\),
with initial condition
\(
\bx_0(0)=[0,0.17,0]^\top
\).
The corresponding reference velocity
\(
\dot{\bx}_0(t)
\)
is used in the formation-error dynamics.

\begin{figure}[h]
\centering
\begin{tikzpicture}[
    >=Stealth,
    scale=0.8,
    every node/.style={font=\small},
    leader/.style={
        rectangle,
        rounded corners=2pt,
        draw=blue!75!black,
        fill=blue!12,
        line width=0.9pt,
        minimum width=9mm,
        minimum height=8mm
    },
    follower/.style={
        circle,
        draw=teal!75!black,
        fill=teal!12,
        line width=0.9pt,
        minimum size=9mm
    },
    pinned/.style={
        ->,
        draw=blue!75!black,
        line width=1.0pt
    },
    communication/.style={
        ->,
        draw=teal!75!black,
        line width=1.0pt
    },
    formation/.style={
        dashed,
        draw=gray!55,
        line width=0.65pt
    }
]

\node[leader] (x0) at (-2.0,0.0) {$\bx_0$};

\node[follower] (x1) at (0.0,-1.0) {$\bx_1$};
\node[follower] (x2) at (1.8,-1.0) {$\bx_2$};
\node[follower] (x3) at (1.8,1.0) {$\bx_3$};
\node[follower] (x4) at (0.0,1.0) {$\bx_4$};

\draw[formation]
    (x1.center) -- (x2.center) --
    (x3.center) -- (x4.center) -- cycle;

\draw[pinned] (x0) -- (x1);

\draw[communication] (x1) -- (x2);
\draw[communication] (x2) -- (x3);
\draw[communication] (x3) -- (x4);

\end{tikzpicture}

\caption{Directed leader--follower communication
topology and prescribed square formation.
Arrows represent information flow, whereas
dashed lines indicate the desired relative
geometry in three-dimensional space.}
\label{fig:topology_formation}
\end{figure}
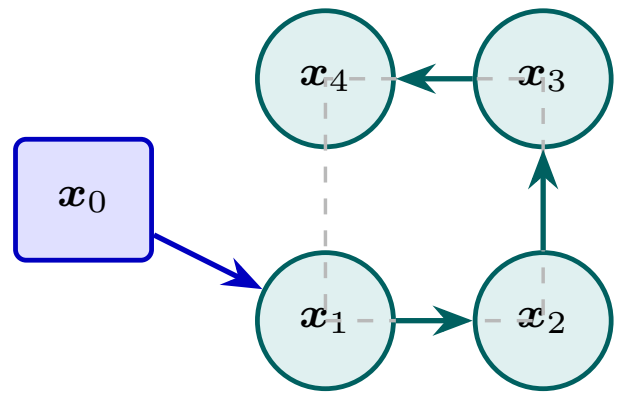

The directed communication topology and prescribed
square formation are illustrated in
Fig.~\ref{fig:topology_formation}.
The communication links satisfy
\(
a_{21}=a_{32}=a_{43}=1
\),
with all other \(a_{ij}=0\), and the
leader-pinning gains are
\(
b_1=1
\)
and
\(
b_2=b_3=b_4=0
\).
The formation offsets are selected as
\(
\bh_1=[-0.4,-0.4,0]^\top
\),
\(
\bh_2=[0.4,-0.4,0]^\top
\),
\(
\bh_3=[0.4,0.4,0]^\top
\),
and
\(
\bh_4=[-0.4,0.4,0]^\top
\).

For each follower, the local augmented
control-weighting matrix is selected as
\(
\bar{\bR}_i=\frac12\bm I_{m_i}
\),
where
\(
m_i=\dim(\bar{\bu}_i)
\).
The prescribed optimal value function is
\(
V_i^*(\be_i)
=
\frac12\be_i^\top\be_i
\),
with the desired fixed-time decay
\(
\dot V_i^*
=
-2(V_i^*)^{3/4}
-
(V_i^*)^{3/2}
\).
Define
\(
\hat F_{V,i}
=
\nabla V_i^{*\top}\hat{\mathbf F}_i^e
\),
\(
\hat A_{e,i}
=
\hat F_{V,i}
+
2(V_i^*)^{3/4}
+
(V_i^*)^{3/2}
\),
and
\(
D_{d,i}
=
\nabla V_i^{*\top}
\mathbf G_i
\bar{\bR}_i^{-1}
\mathbf G_i^\top
\nabla V_i^*
\).
For \(\be_i\neq\bm0\), by Theorem~\ref{theorem2}, the inverse-optimal
running-cost components are selected as
\(
\mathbf L_{1i}^\top
=
(2\hat A_{e,i}/D_{d,i}-1)
\nabla V_i^{*\top}\mathbf G_i
\)
and
\(
L_{2i}
=
-\hat F_{V,i}
+
\hat A_{e,i}^2/D_{d,i}
\). During online learning, the behavior input
consists of the current critic-induced policy
and a bounded probing signal,
\(
\bar{\bu}_i(t)
=
\bar{\bu}_i(t)+\bar{\mathbf e}_{u_i}(t)
\).
The probing signal is used to collect informative
integral Bellman data, while the critic weights
are adapted using current data and experience
replay according to \eqref{eq:critic_update}.
The probing amplitude is subsequently reduced
after sufficient replay informativity is obtained. The reason a probing signal is used in the simulation setup is to increase the value of \(\alpha_i\) in~\eqref{eq:replay_ie}.
For all \(i\in\mathcal V\). For each follower, the critic basis is selected as
\(
\boldsymbol\phi_i(\be_i)
=
[e_{i1}^2,e_{i2}^2,e_{i3}^2,
e_{i1}e_{i2},e_{i1}e_{i3},e_{i2}e_{i3},
e_{i1}^4,e_{i2}^4,e_{i3}^4]^\top
\in\mathbb R^9
\).
The critic weights are initialized as
\(
\hat{\bW}_i(0)=\bm0_{9}
\),
for
\(
i=1,\ldots,4
\).
The corresponding ideal weights are
\(
\bW_i^*
=
[0.5,0.5,0.5,0,0,0,0,0,0]^\top
\),
consistent with
\(
V_i^*(\be_i)=\frac12\be_i^\top\be_i
\). 

\subsection{Simulation Results}

The simulation results are presented to evaluate the leader–follower
formation-tracking performance and the fixed-time convergence be
havior of the proposed distributed IRL framework.

\begin{figure}[H]
    \centering
    \scalebox{1}[0.8]{%
        \includegraphics[width=\columnwidth]{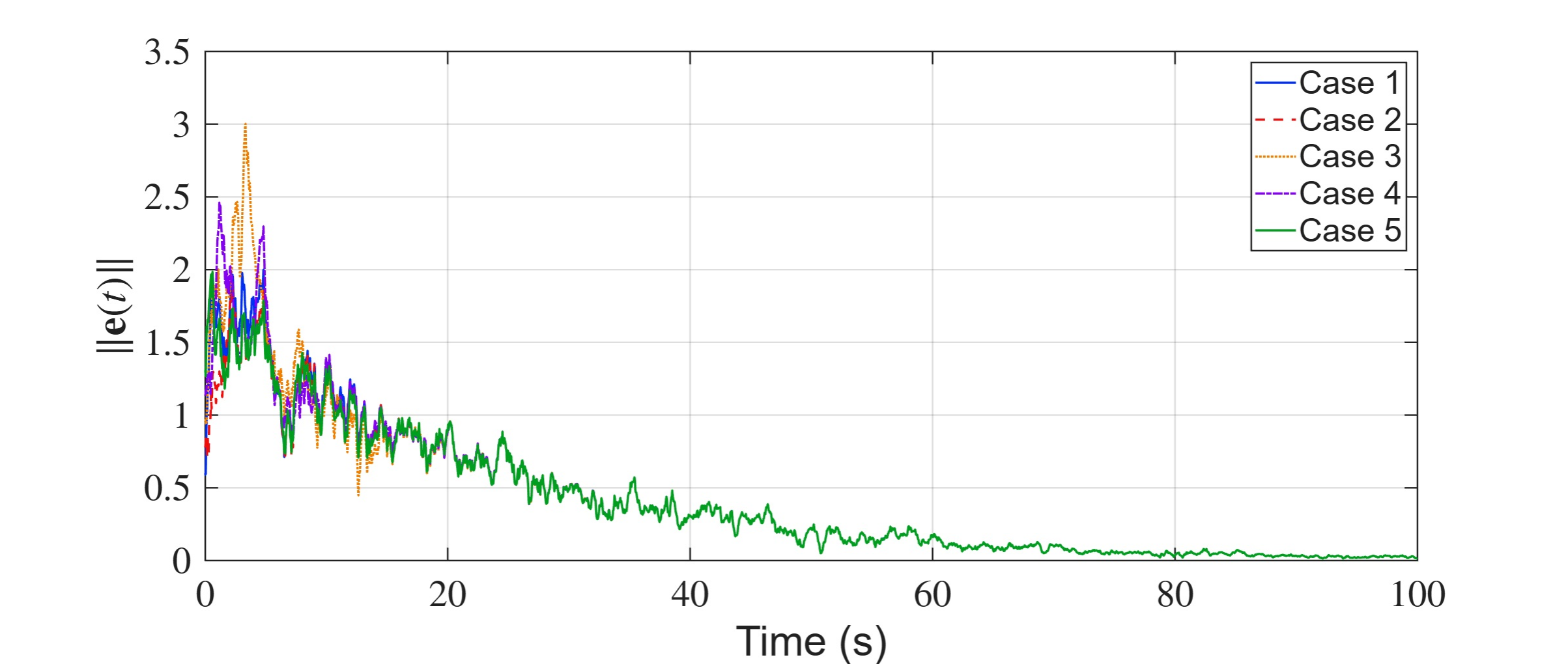}%
    }
    \caption{Overall leader--follower formation-tracking error
    under five initial conditions:
    \(\|\mathbf e^{(s)}(0)\|
    =\{0.6,1.0,1.4,1.8,2.2\}\),
    \(s=1,\ldots,5\), where \(s\) denotes the
    initial-condition case index.}
    \label{fig:N20_1}
\end{figure}

\begin{figure}[H]
    \centering
    \scalebox{1}[0.8]{%
        \includegraphics[width=\columnwidth]{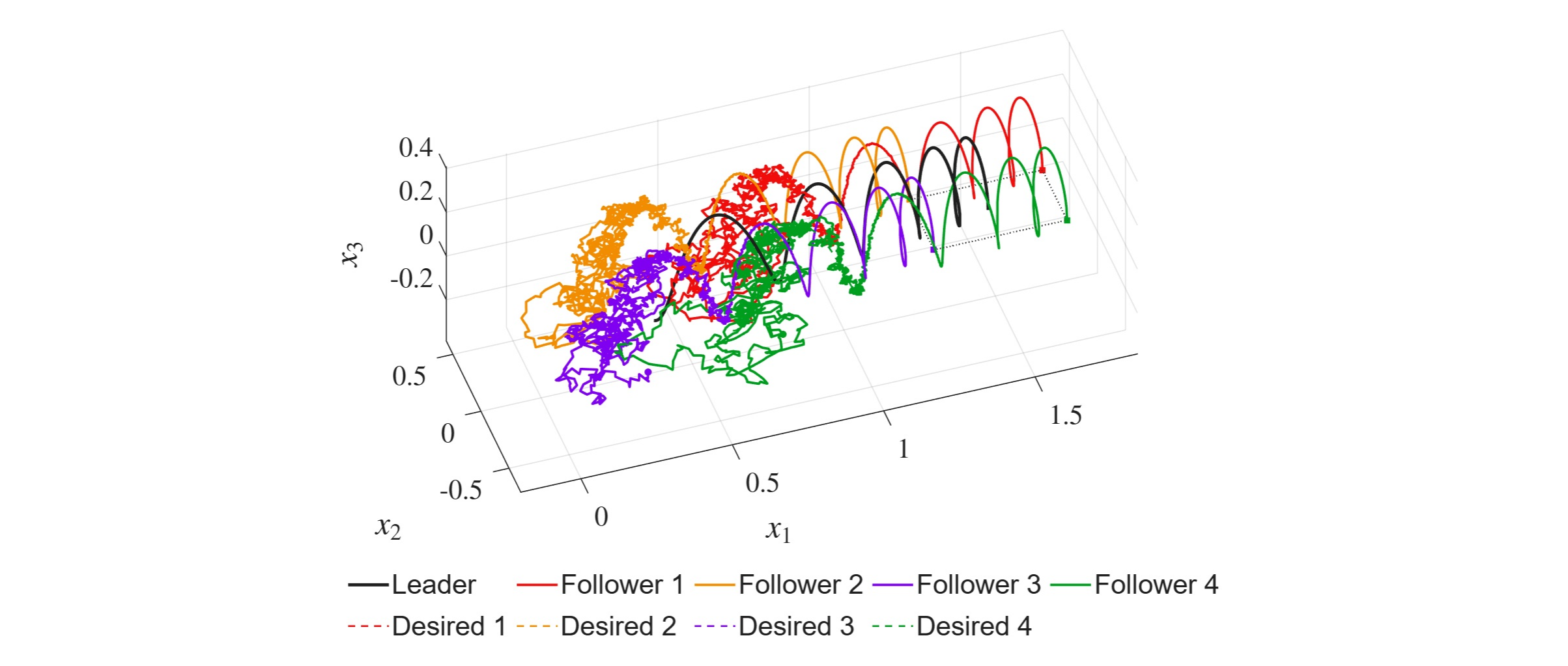}%
    }
    \caption{Three-dimensional leader--follower
    formation-tracking trajectories along the
    prescribed reference path. The initial phase is a period of strong stimulation aimed at gathering a sufficient amount of informative data; consequently, the trajectories of the agents are relatively chaotic.}
    \label{fig:N20_11}
\end{figure}

\begin{figure}[H]
    \centering
    \scalebox{1}[0.8]{%
        \includegraphics[width=\columnwidth]{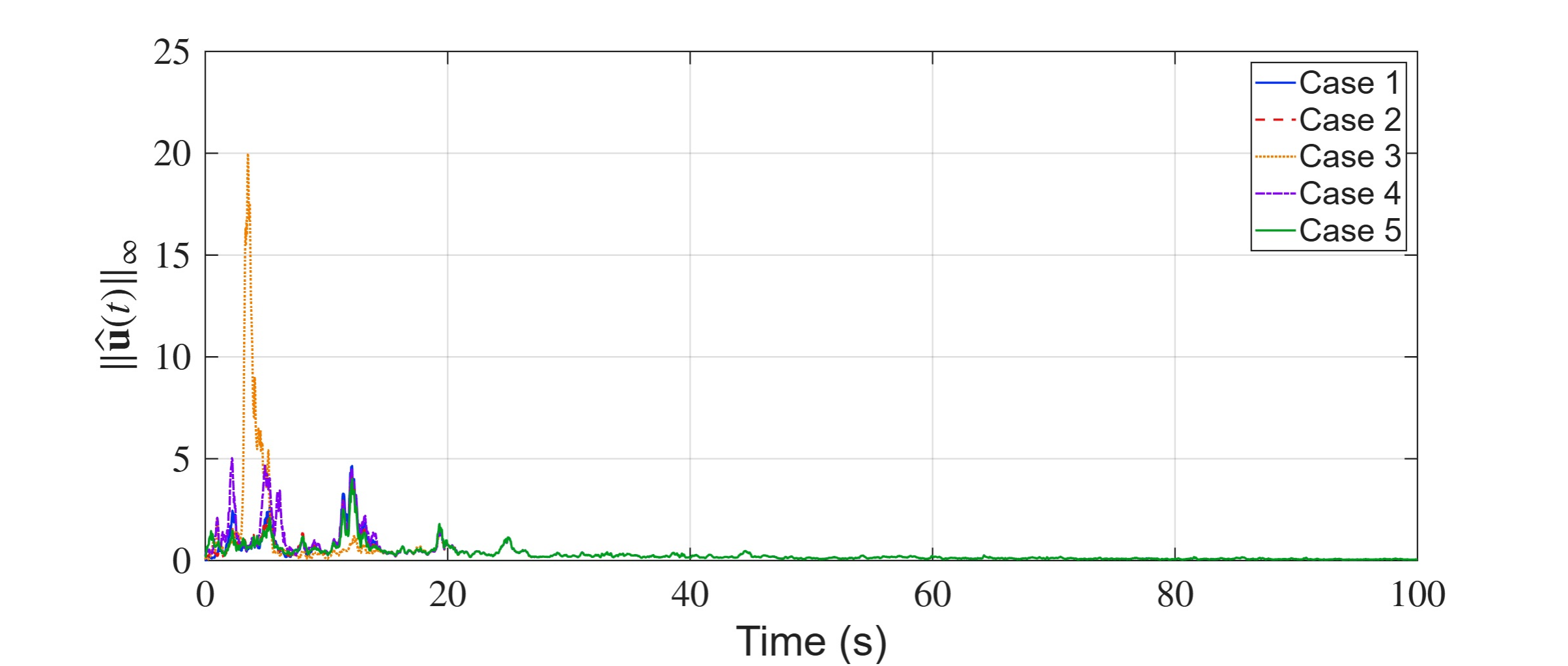}%
    }
    \caption{Evolution of
    \(\|\hat{\mathbf{u}}(t)\|_{\infty}
    =\max_{i=1,\ldots,4}
    \|\hat{\bar{\bu}}_i(\be_i(t))\|_{\infty}
    =\max_{i=1,\ldots,4}\max_{j=1,2,3}
    |\hat{\bar u}_{i,j}(\be_i(t))|\)
    under five different initial conditions.}
    \label{fig:N20_4}
\end{figure}

\begin{figure}[H]
    \centering
    \scalebox{1}[0.8]{%
        \includegraphics[width=\columnwidth]{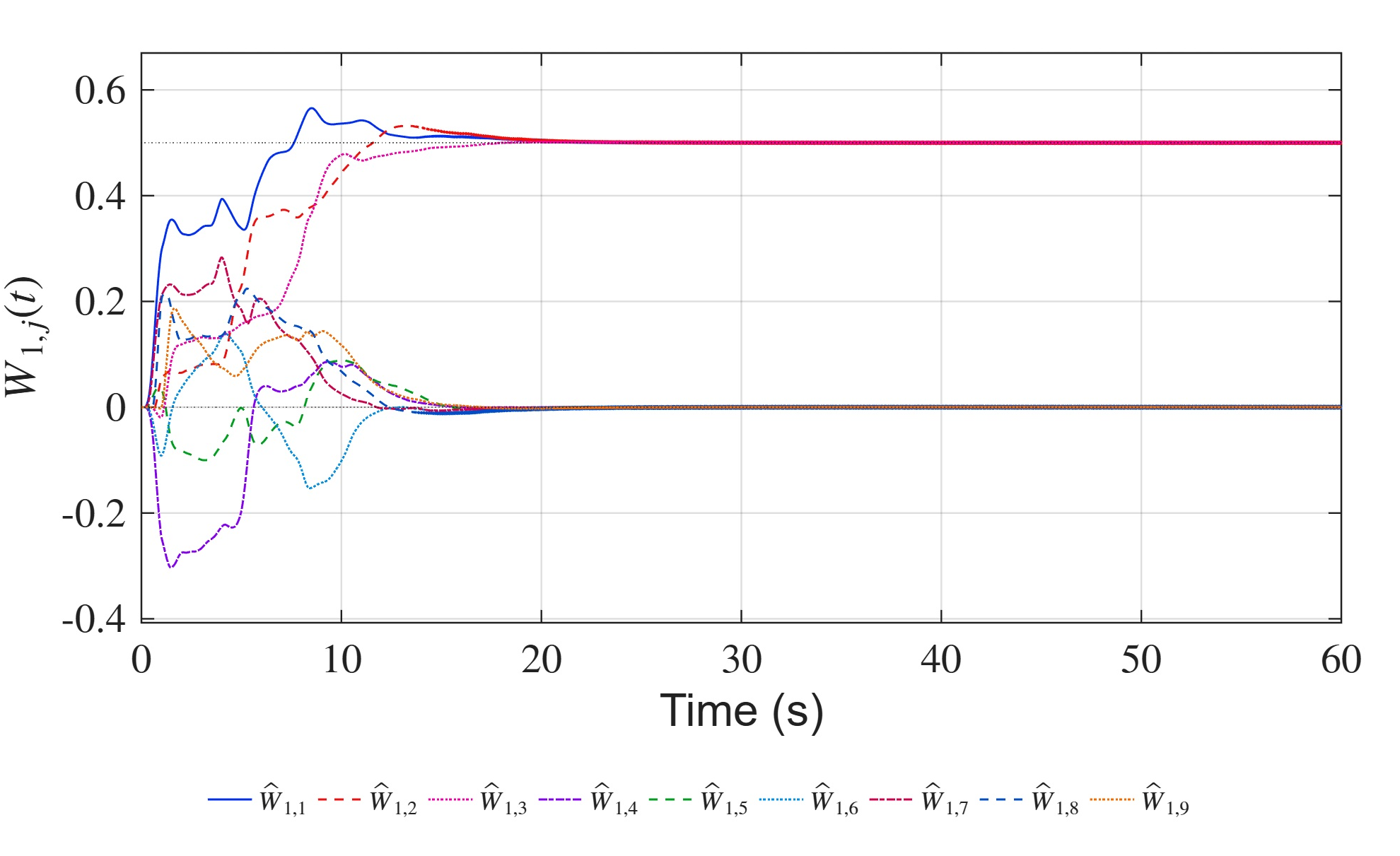}%
    }
    \caption{Evolution of the critic weights
    of follower 1 under Case 1:
    \(\|\be^{(1)}(0)\|=0.6\),
    \(\hat{\mathbf W}_1(0)=\mathbf 0_9\).}
    \label{fig:N20_6}
\end{figure}

\begin{figure}[H]
    \centering
    \scalebox{1}[0.95]{%
        \includegraphics[width=\columnwidth]{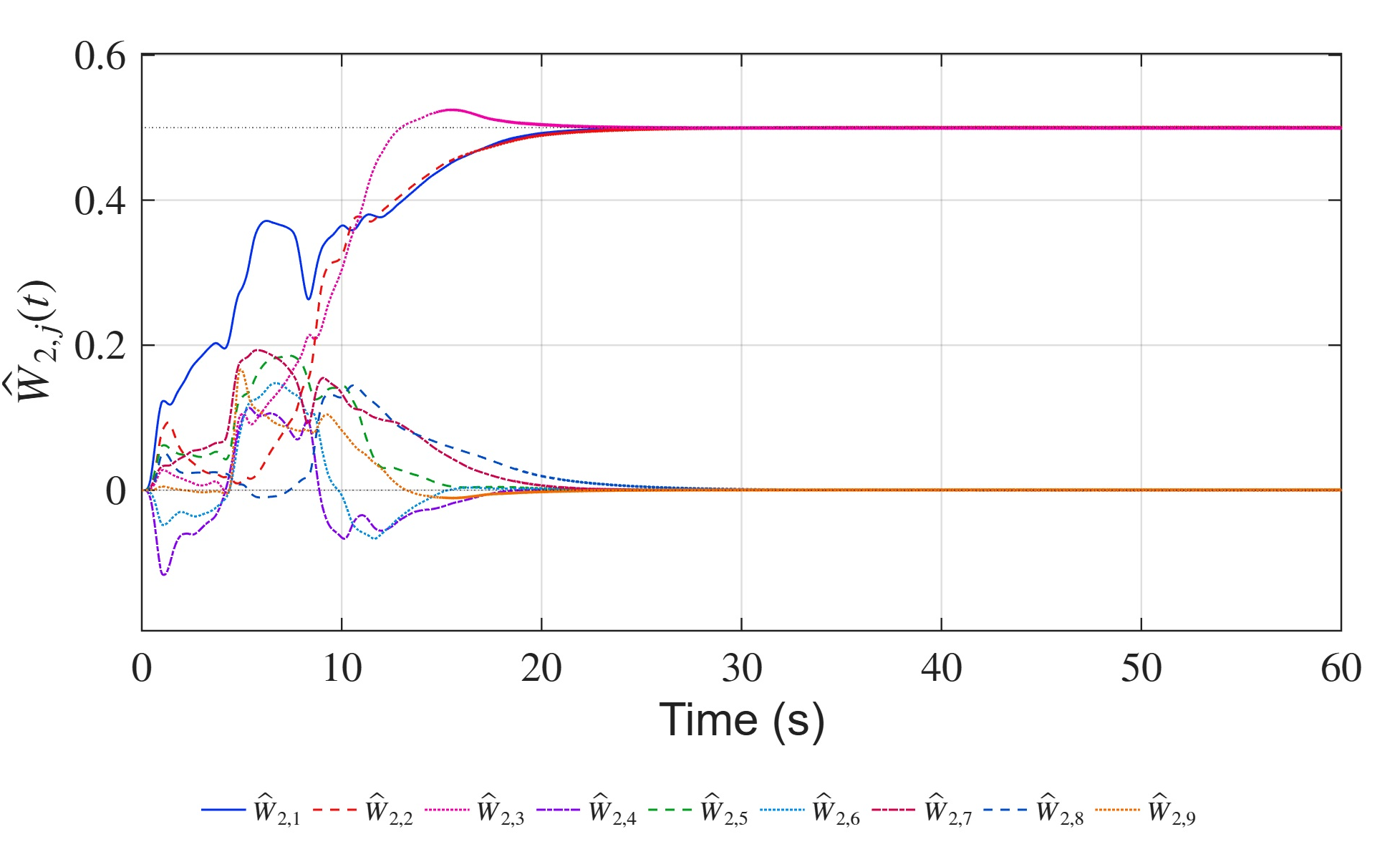}%
    }
    \caption{Evolution of the critic weights
    of follower 2 under Case 1:
    \(\|\be^{(1)}(0)\|=0.6\),
    \(\hat{\mathbf W}_2(0)=\mathbf 0_9\).}
    \label{fig:N20_7}
\end{figure}

\begin{figure}[H]
    \centering
    \scalebox{1}[0.85]{%
        \includegraphics[width=\columnwidth]{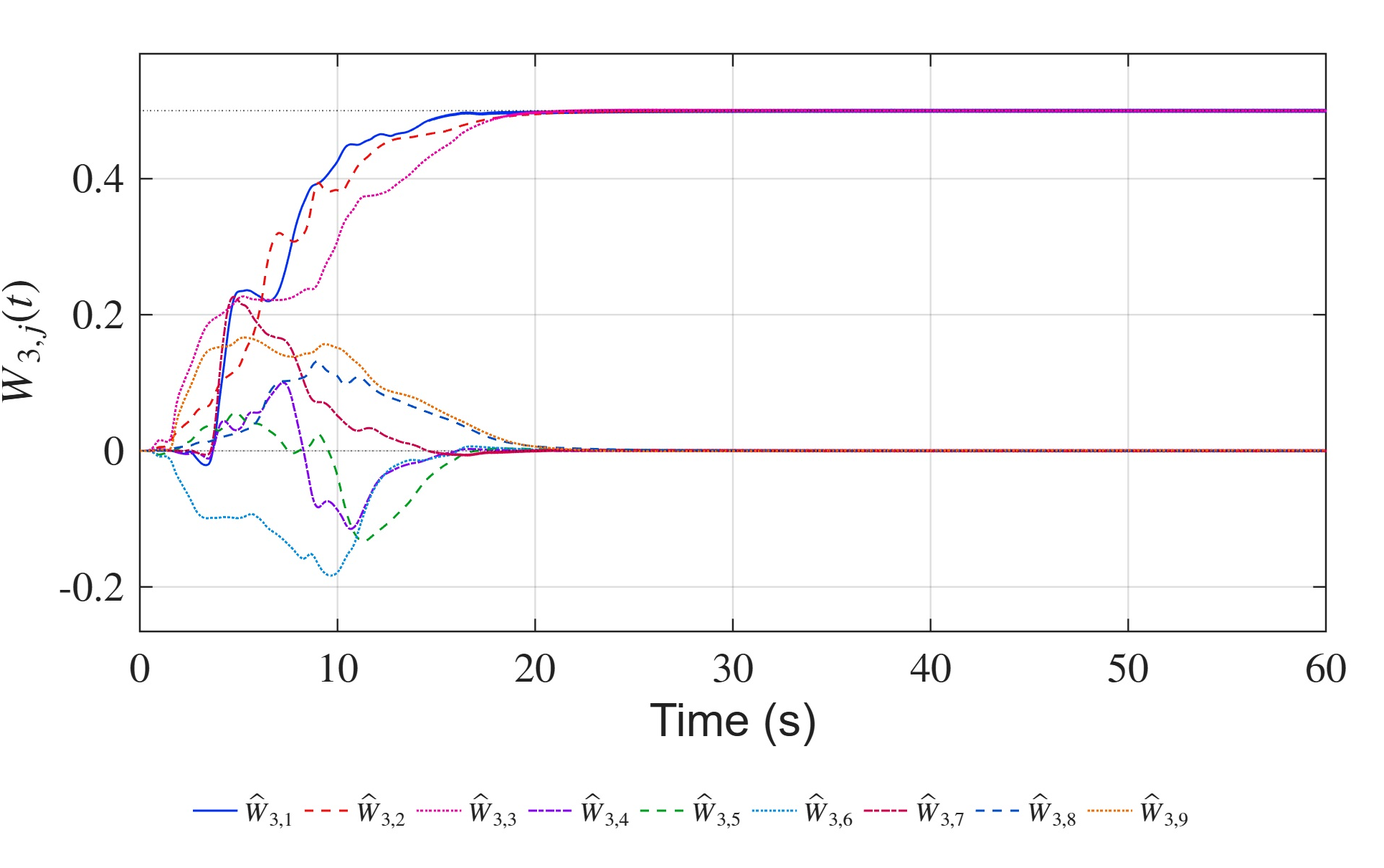}%
    }
    \caption{Evolution of the critic weights
    of follower 3 under Case 1:
    \(\|\be^{(1)}(0)\|=0.6\),
    \(\hat{\mathbf W}_3(0)=\mathbf 0_9\).}
    \label{fig:N20_8}
\end{figure}

\begin{figure}[H]
    \centering
    \scalebox{1}[1.0]{%
        \includegraphics[width=\columnwidth]{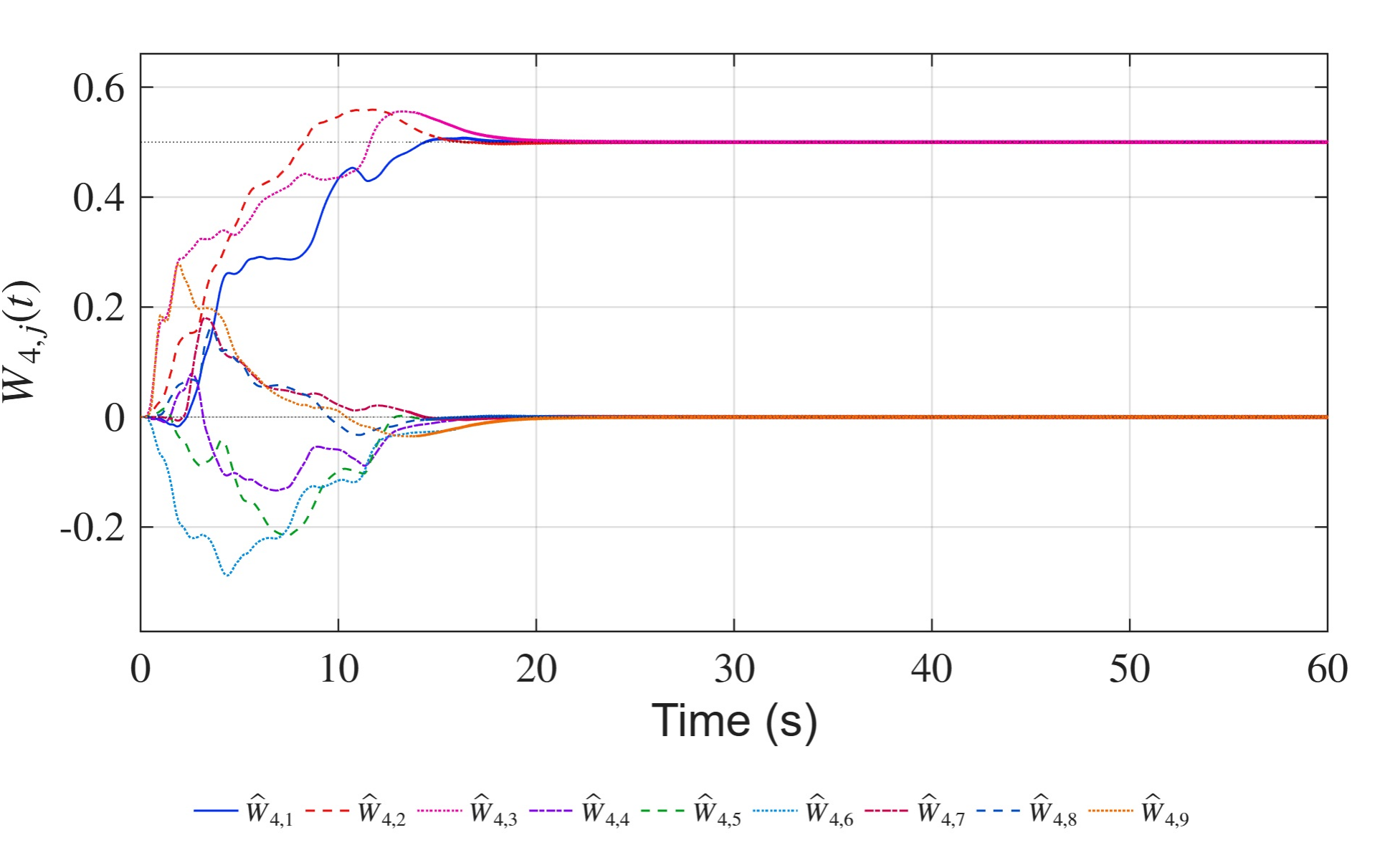}%
    }
    \caption{Evolution of the critic weights
    of follower 4 under Case 1:
    \(\|\be^{(1)}(0)\|=0.6\),
    \(\hat{\mathbf W}_4(0)=\mathbf 0_9\).}
    \label{fig:N20_9}
\end{figure}

\begin{figure}[H]
    \centering
    \scalebox{1}[1.0]{%
        \includegraphics[width=\columnwidth]{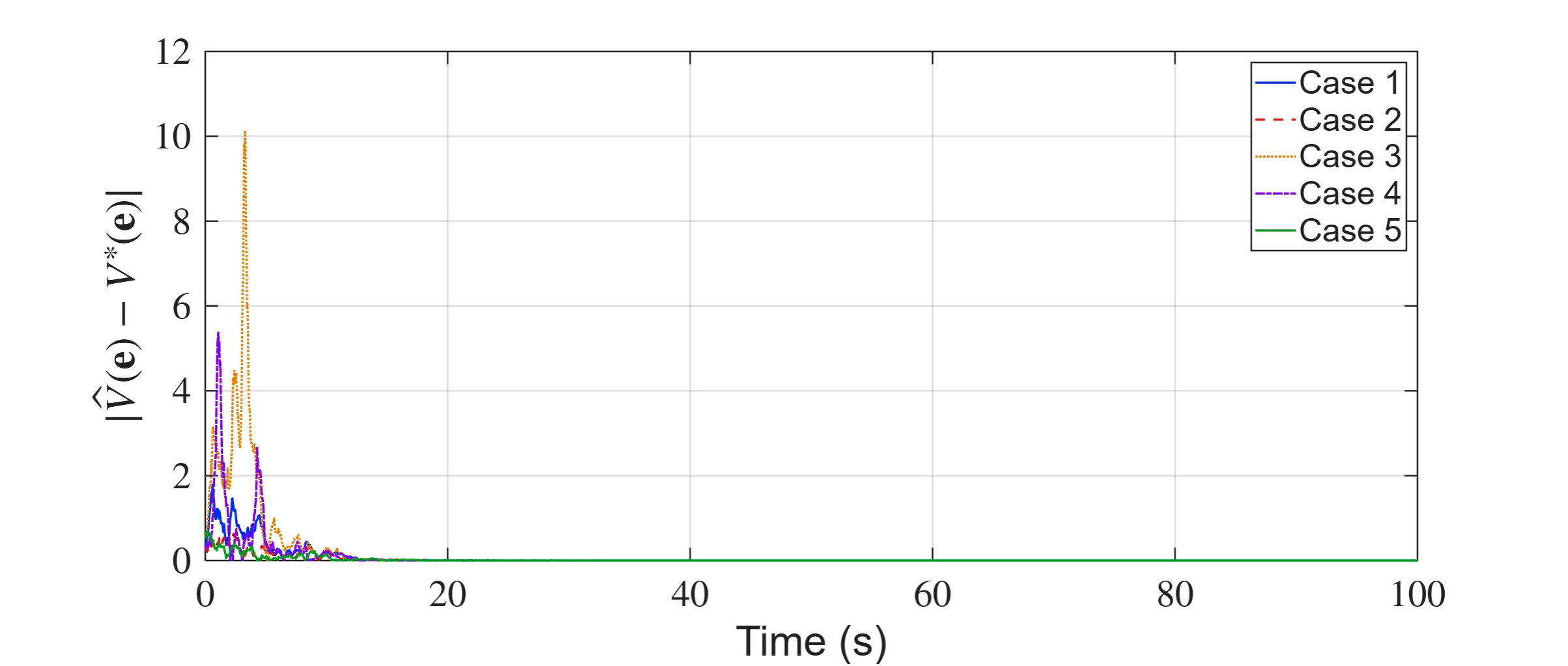}%
    }
    \caption{Evolution of the critic value-function
    approximation error under five different
    initial conditions.}
    \label{fig:N20_10}
\end{figure}

\begin{figure}[H]
    \centering
    \scalebox{1}[1.0]{%
        \includegraphics[width=\columnwidth]{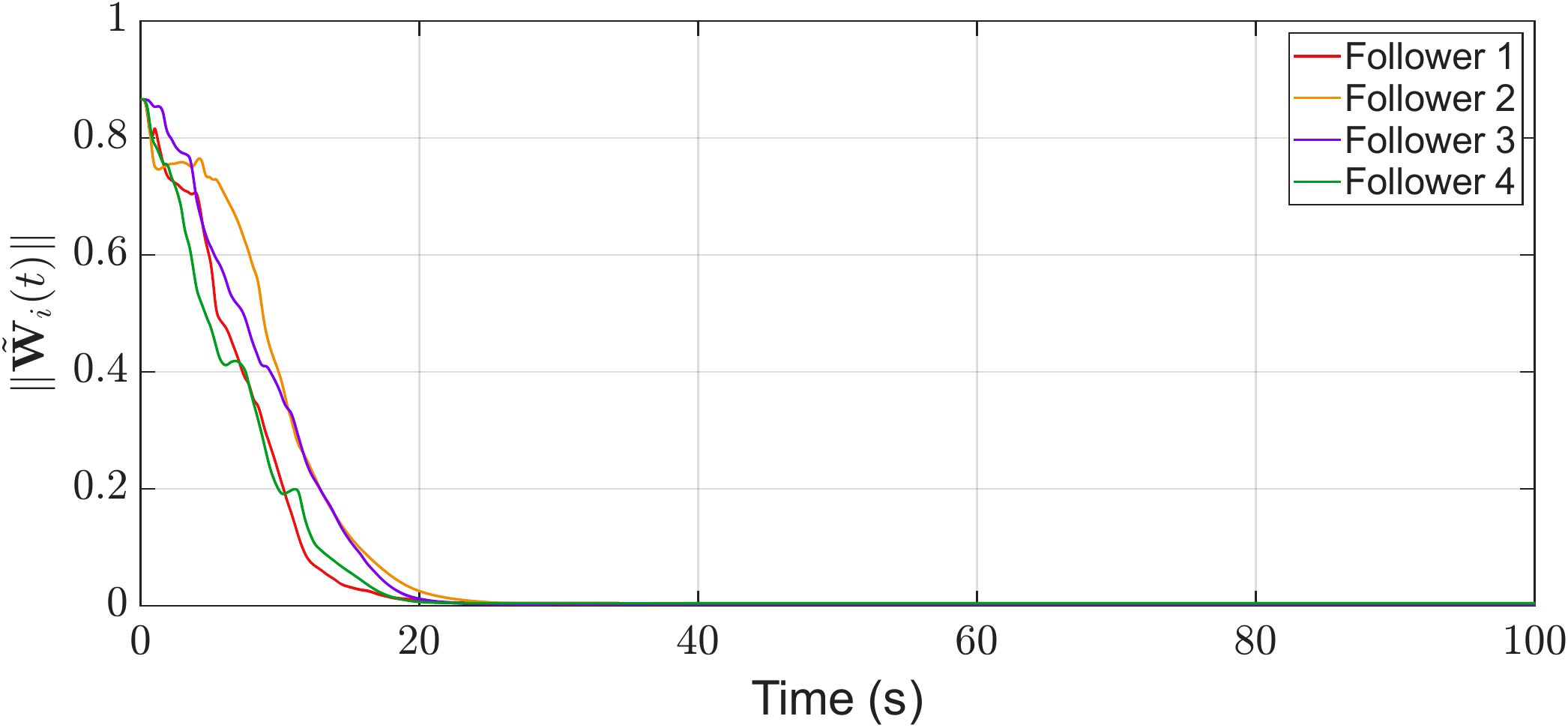}%
    }
    \caption{Evolution of the critic-weight estimation-error
    norms of the four followers under Case 1:
    \(\hat{\mathbf W}_i(0)=\mathbf 0_9\),
    \(i=1,\ldots,4\).}
    \label{fig:N20_15}
\end{figure}

\begin{figure}[H]
    \centering
    \scalebox{1}[1.0]{%
        \includegraphics[width=\columnwidth]{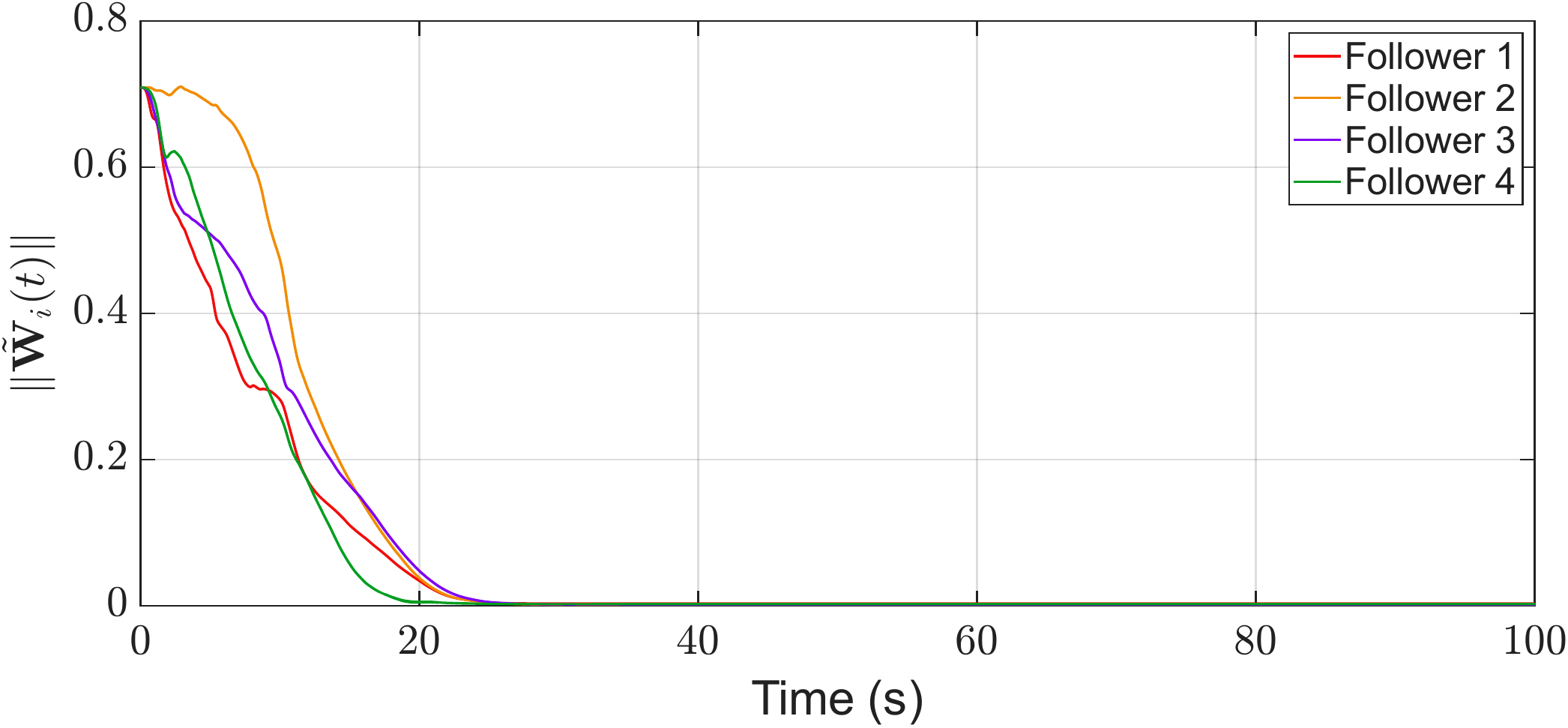}%
    }
    \caption{Evolution of the critic-weight estimation-error
    norms of the four followers under Case 2:
    \(\hat{\mathbf W}_i(0)=0.15\mathbf 1_9\),
    \(i=1,\ldots,4\).}
    \label{fig:N20_19}
\end{figure}

\begin{figure}[H]
    \centering
    \scalebox{1}[0.7]{%
        \includegraphics[width=\columnwidth]{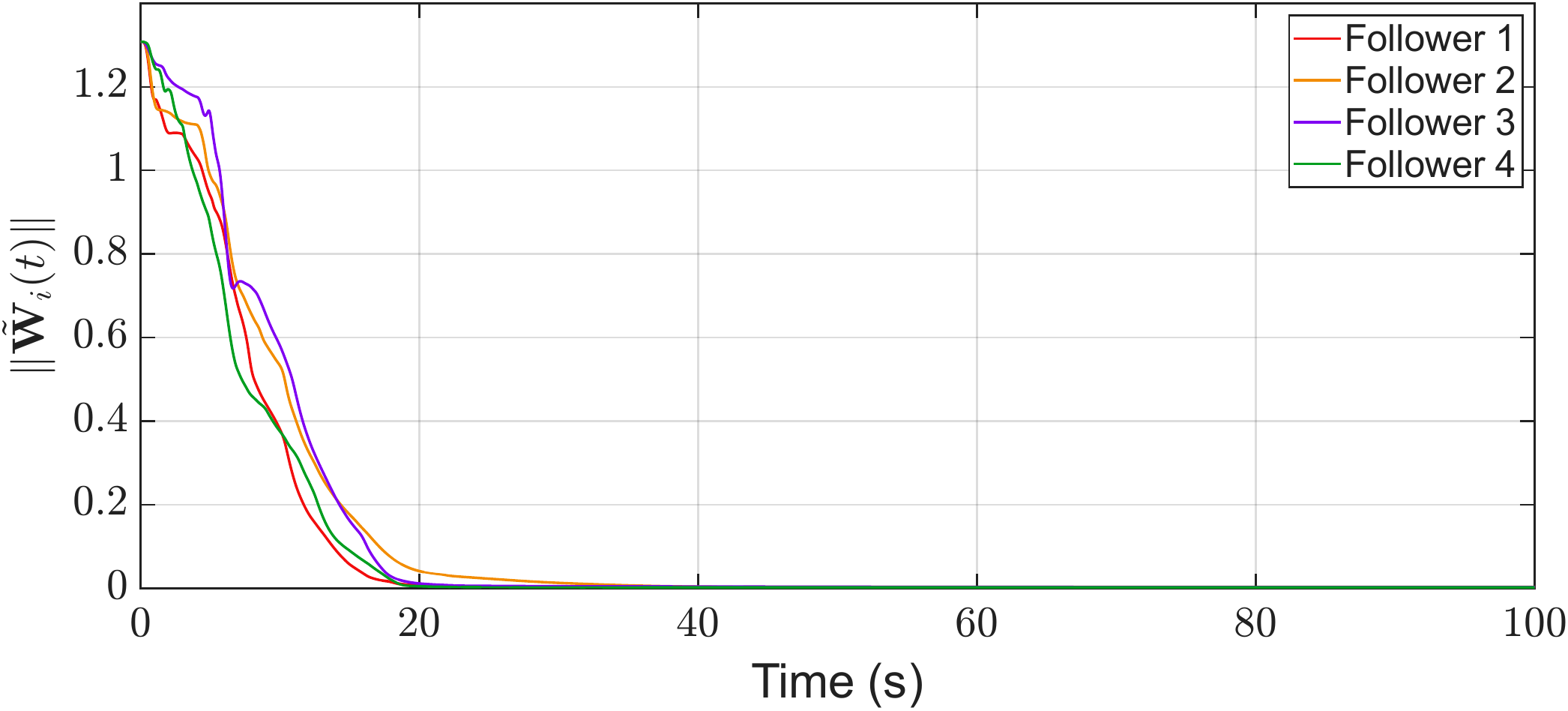}%
    }
    \caption{Evolution of the critic-weight estimation-error
    norms of the four followers under Case 3:
    \(\hat{\mathbf W}_i(0)=-0.20\mathbf 1_9\),
    \(i=1,\ldots,4\).}
    \label{fig:N20_20}
\end{figure}

Fig.~\ref{fig:N20_1} shows that, despite the five distinct
initial formation-error norms ranging from \(0.6\) to \(2.2\),
all trajectories exhibit a common decreasing trend after
the initial probing-induced transients. The largest
transient error is approximately \(3.0\), whereas the
formation-error norms fall below approximately \(0.5\)
after \(30\,\mathrm{s}\) and approach a small neighborhood
of the origin by \(100\,\mathrm{s}\).
The three-dimensional trajectories in
Fig.~\ref{fig:N20_11} further illustrate the recovery
of the prescribed leader--follower formation along
the reference path.

As shown in Fig.~\ref{fig:N20_4}, the learned control
magnitude reaches a transient peak of approximately
\(20\) in Case 3, while the remaining cases exhibit
considerably smaller peaks. The control magnitudes
subsequently decrease toward zero, indicating that
the large control actions are primarily confined
to the initial learning transient. This is entirely consistent with the argument in remark \ref{rmk3}.

Figs.~\ref{fig:N20_6}--\ref{fig:N20_9} show that the
first three critic weights of each follower converge
toward \(0.5\), whereas the remaining six weights
approach zero. The weight trajectories settle near
their prescribed values within approximately
\(20\)--\(25\,\mathrm{s}\), despite temporary overshoots
and sign changes. Consistently,
Fig.~\ref{fig:N20_10} shows that the value-function
approximation error, whose largest transient peak
is approximately \(10\), rapidly decreases to a
negligible level after roughly \(15\,\mathrm{s}\).

Finally, Figs.~\ref{fig:N20_15},
\ref{fig:N20_19}, and \ref{fig:N20_20} demonstrate
critic-weight error reduction under three distinct
critic initializations. Although the initial
per-follower error norms differ across these
experiments, reaching approximately \(0.9\),
\(0.7\), and \(1.3\), respectively, all four
followers attain nearly zero estimation errors
within approximately \(20\)--\(25\,\mathrm{s}\).
These observations support the effectiveness
of the integral Bellman learning and experience
replay mechanisms and are consistent with
practical fixed-time convergence over the
tested initial conditions.





\section{Conclusion}
\label{sec:conclusion}

This paper developed a distributed inverse-optimal fixed-time
integral reinforcement learning framework for unknown nonlinear
leader--follower multi-agent systems. Unknown agent dynamics were
reconstructed from finite integral data, and an inverse-optimal
cost was constructed from a prescribed fixed-time value function.
A critic-only integral reinforcement learning scheme with finite
replay informativity was developed, and practical fixed-time
convergence of both critic-weight errors and leader-following
consensus errors was established. Beyond the second-order
multi-agent setting considered in \cite{Gong2025}, the proposed
framework accommodates unknown nonlinear agent dynamics.
Furthermore, it extends the single-system optimal stabilization
frameworks of \cite{Kokolakis2023FixedTime,Kokolakis2025Predefined}
to distributed leader--follower control through integral
identification and inverse-optimal synthesis.
Future work may consider communication constraints,
disturbances, input saturation, and time-varying network
topologies.

\section*{Acknowledgment}
The authors would like to thank Ho Chi Minh City University of Technology (HCMUT) and Vietnam National University Ho Chi Minh City (VNU-HCM) for supporting this research.

\appendices

\section{Proof of lemma~\ref{lem:derivative_limited_window_bound}}
\label{Movingwin}
\begin{proof}
Let
\(
M_y:=\max_{\tau\in[t-T,t]}y(\tau)
\)
and choose
\(
\tau_m\in[t-T,t]
\)
such that
\(
y(\tau_m)=M_y
\).
Since at least one of the intervals
\(
[t-T,\tau_m]
\)
and
\(
[\tau_m,t]
\)
has length no smaller than \(T/2\), and
\(
|\dot y(\tau)|\leq L_y
\)
almost everywhere, absolute continuity gives
\(
|y(\tau)-y(\tau_m)|
\leq
\int_{\min\{\tau,\tau_m\}}^{\max\{\tau,\tau_m\}}
|\dot y(s)|\,ds
\leq
L_y|\tau-\tau_m|
\).
Thus,
\(
y(\tau)\geq M_y-L_y|\tau-\tau_m|
\),
and hence
\(
y(\tau)\geq M_y/2
\)
whenever
\(
|\tau-\tau_m|\leq M_y/(2L_y)
\).
Consequently, there exists a one-sided interval
of length
\(
\ell_y:=\min\{T/2,M_y/(2L_y)\}
\)
on which
\(
y(\tau)\geq M_y/2
\).
Therefore,
\begin{equation}
\int_{t-T}^{t}y(\tau)\,d\tau
\geq
\frac{M_y}{2}
\min\left\{
\frac{T}{2},
\frac{M_y}{2L_y}
\right\}.
\label{eq:window_area_lower_bound}
\end{equation}

Let
\(
I_y:=\int_{t-T}^{t}y(\tau)\,d\tau
\).
For
\(
0<\gamma<1
\),
the function
\(
s^{\gamma-1}
\)
is decreasing on \(s>0\).
Since
\(
0\leq y(\tau)\leq M_y
\),
one obtains
\(
y^\gamma(\tau)
=
y(\tau)y^{\gamma-1}(\tau)
\geq
M_y^{\gamma-1}y(\tau)
\),
with the inequality extended to \(y(\tau)=0\)
by continuity.
Integrating yields
\(
\int_{t-T}^{t}y^\gamma(\tau)\,d\tau
\geq M_y^{\gamma-1}I_y
\).

First, if
\(
M_y\leq L_yT
\),
then
\eqref{eq:window_area_lower_bound}
implies
\(
I_y\geq M_y^2/(4L_y)
\),
so that
\(
M_y\leq2(L_yI_y)^{1/2}
\).
Since
\(
\gamma-1<0
\),
raising this bound to the power
\(
\gamma-1
\)
reverses the inequality, giving
\(
M_y^{\gamma-1}
\geq
2^{\gamma-1}
L_y^{\frac{\gamma-1}{2}}
I_y^{\frac{\gamma-1}{2}}
\).
Consequently,
\(
\int_{t-T}^{t}y^\gamma(\tau)\,d\tau
\geq
2^{\gamma-1}
L_y^{\frac{\gamma-1}{2}}
I_y^{1+\frac{\gamma-1}{2}}
\).
Since
\(
1+\frac{\gamma-1}{2}
=
\frac{\gamma+1}{2}
=
\mu_\gamma
\),
it follows that
\begin{equation}
\int_{t-T}^{t}y^\gamma(\tau)\,d\tau
\geq
2^{\gamma-1}
L_y^{\frac{\gamma-1}{2}}
I_y^{\mu_\gamma}.
\label{eq:window_case_1}
\end{equation}

Next, if
\(
M_y>L_yT
\),
then
\eqref{eq:window_area_lower_bound}
gives
\(
I_y\geq M_yT/4
\),
and hence
\(
M_y\leq4I_y/T
\).
Because
\(
\gamma-1<0
\),
one has
\(
M_y^{\gamma-1}
\geq
(4I_y/T)^{\gamma-1}
=
4^{\gamma-1}T^{1-\gamma}I_y^{\gamma-1}
\).
Therefore,
\(
\int_{t-T}^{t}y^\gamma(\tau)\,d\tau
\geq
4^{\gamma-1}T^{1-\gamma}I_y^\gamma
\).
Moreover,
\(
\gamma-\mu_\gamma
=
(\gamma-1)/2<0
\).
Using
\(
I_y\leq\bar I
\)
gives
\(
I_y^{\gamma-\mu_\gamma}
\geq
\bar I^{\frac{\gamma-1}{2}}
\),
and hence
\(
I_y^\gamma
=
I_y^{\mu_\gamma}
I_y^{\gamma-\mu_\gamma}
\geq
\bar I^{\frac{\gamma-1}{2}}
I_y^{\mu_\gamma}
\).
Thus,
\begin{equation}
\int_{t-T}^{t}y^\gamma(\tau)\,d\tau
\geq
4^{\gamma-1}
T^{1-\gamma}
\bar I^{\frac{\gamma-1}{2}}
I_y^{\mu_\gamma}.
\label{eq:window_case_2}
\end{equation}

Combining
\eqref{eq:window_case_1}
and
\eqref{eq:window_case_2}
yields
\(
\int_{t-T}^{t}y^\gamma(\tau)\,d\tau
\geq
\min\left\{
2^{\gamma-1}L_y^{\frac{\gamma-1}{2}},
4^{\gamma-1}T^{1-\gamma}
\bar I^{\frac{\gamma-1}{2}}
\right\}
I_y^{\mu_\gamma}
\).
Substituting
\(
I_y=\int_{t-T}^{t}y(\tau)\,d\tau
\)
establishes
\eqref{eq:window_low_power_lemma}.
\end{proof}

\section{Proof of Lemma~\ref{lem:critic_coupling_absorption}}
\label{Appen1}

\begin{proof}
For arbitrary
\(
\varepsilon_i>0
\),
consider
\(
\psi_i(s)
:=
c_{c_i}s-\varepsilon_i s^{2\mu_i}
\),
\(s\geq0\).
Since
\(
1/2<\mu_i<1
\),
one has
\(
2\mu_i-1>0
\).
Differentiating yields
\(
\psi_i'(s)
=
c_{c_i}
-
2\mu_i\varepsilon_i s^{2\mu_i-1}
\).
For
\(
c_{c_i}>0
\),
the stationary condition
\(
\psi_i'(s)=0
\)
gives
\(
s_i^{*}
=
\left(
\frac{c_{c_i}}
{2\mu_i\varepsilon_i}
\right)^{\frac{1}{2\mu_i-1}}
\).
Moreover,
\(
\psi_i''(s)
=
-2\mu_i(2\mu_i-1)
\varepsilon_i s^{2\mu_i-2}
<0
\)
for \(s>0\), so
\(
s_i^{*}
\)
is the unique global maximizer of
\(
\psi_i
\)
on \(s\geq0\).

At
\(
s=s_i^{*}
\),
the stationary condition implies
\(
\varepsilon_i(s_i^{*})^{2\mu_i}
=
c_{c_i}s_i^{*}/(2\mu_i)
\).
Consequently,
\(
\begin{aligned}
\psi_i(s_i^{*})
&=
c_{c_i}s_i^{*}
-
\frac{c_{c_i}s_i^{*}}{2\mu_i}
=
\frac{2\mu_i-1}{2\mu_i}
c_{c_i}
\left(
\frac{c_{c_i}}
{2\mu_i\varepsilon_i}
\right)^{\frac{1}{2\mu_i-1}}
=
\Delta_{c_i}.
\end{aligned}
\)

Since
\(
\psi_i(s)\leq\psi_i(s_i^{*})
\)
for all \(s\geq0\), it follows that
\(
c_{c_i}s-\varepsilon_i s^{2\mu_i}
\leq\Delta_{c_i}
\),
or equivalently,
\(
c_{c_i}s
\leq
\varepsilon_i s^{2\mu_i}
+\Delta_{c_i}
\).
For
\(
c_{c_i}=0
\),
the same inequality holds with
\(
\Delta_{c_i}=0
\).
This completes the proof.
\end{proof}

\bibliographystyle{IEEEtran}
\bibliography{references}

\raggedbottom

\makeatletter
\def\@IEEEBIOskipN{0.5\baselineskip}
\makeatother

\vspace{0.3em}
\vspace{-0.6em}

\vspace{0.5em}

\end{document}